\documentclass[%
 aip,
 amsmath,amssymb,
 reprint,%
 twocolumn
]{revtex4-1}

\usepackage{graphicx}% Include figure files
\usepackage{dcolumn}% Align table columns on decimal point
\usepackage{bm}% bold math
\usepackage[colorlinks,linkcolor=red,citecolor=blue,urlcolor=blue]{hyperref}

\usepackage[utf8]{inputenc}
\usepackage[T1]{fontenc}
\usepackage{mathptmx}
\usepackage{subcaption}
\usepackage{float}
\usepackage{stmaryrd}
\usepackage{tabularx}
\usepackage{booktabs}

\usepackage{soul,color}

\allowdisplaybreaks

\newcommand{\rcol}[1]{%
  \begingroup
    \textcolor{black}{#1}%
  \endgroup
}

\begin{document}

\preprint{AIP/123-QED}

%\title{Conservative and efficient derivation of adaptive lattice Boltzmann methods for the simulation of compressible flows}

%\title{Interpolation-free formulation of adaptive velocity discretizations for compressible lattice Boltzmann methods}

\title{\rcol{Compressible lattice Boltzmann methods with adaptive velocity stencils: An interpolation-free formulation}}

%\title{Adaptive velocity stencils for compressible lattice Boltzmann \rcol{without interpolation}}
%\title{Adaptive velocity stencils for compressible lattice Boltzmann with on-grid propagation}
% !!! proposition de titre: ... with on-lattice propagation.
\author{C. Coreixas} \email[Corresponding author: ]{christophe.coreixas@unige.ch}
\affiliation{Department of Computer Science, University of Geneva, 1204 Geneva, Switzerland}%

%\author{S. A. Hosseini}%
 %\email{seyed.hosseini@ovgu.de}
%	\affiliation{Laboratory of Fluid Dynamics and Technical Flows, University of Magdeburg ``Otto von Guericke'', D-39106 Magdeburg, Germany}%
%	\affiliation{Laboratoire EM2C, CNRS, CentraleSup\'elec, Universit\'e   Paris-Saclay, 3 rue Joliot Curie, 91192, Gif-sur-Yvette Cedex, France}%
%	\affiliation{International Max Planck Research School (IMPRS) for Advanced Methods in Process and Systems Engineering, Magdeburg, Germany}

\author{J. Latt}
 \email{jonas.latt@unige.ch}
\affiliation{Department of Computer Science, University of Geneva, 1204 Geneva, Switzerland}%

\date{\today}% It is always \today, today,

\begin{abstract}

Adaptive lattice Boltzmann methods (LBMs) are based on velocity discretizations that self-adjust to local macroscopic conditions such as velocity and temperature. While this feature improves the accuracy and the stability of LBMs for large velocity and temperature fluctuations, it also strongly impacts the efficiency of the algorithm due to space interpolations that are required to get populations at grid nodes. 
To avoid this defect, the present work proposes new formulations of adaptive LBMs which do not rely anymore on space interpolations, hence, drastically improving their parallel efficiency for the simulation of high-speed compressible flows.
To reach this goal, the adaptive phase discretization is restricted to particular states that are compliant with the efficient ``collide and stream'' algorithm, and as a consequence, it does not require additional interpolation steps.
The development of proper state-adaptive solvers with on-grid propagation imposes new restrictions and challenges on the discrete stencils, namely the need for an extended operability range allowing for the transition between two phase discretizations. 
Achieving the minimum operability range for discrete polynomial equilibria requires rather large stencils (e.g. D2Q81, D2Q121) and is therefore not competitive for compressible flow simulations. 
However, as shown in the article, the use of numerical equilibria can provide for overlaps in the operability ranges of neighboring discrete shifts at acceptable cost using the D2Q21 lattice. Through several numerical validations, the present approach is shown to allow for an efficient realization of discrete state-adaptive LBMs for high Mach number flows even in the low viscosity regime.
\end{abstract}

\maketitle

\section{\label{sec:introduction}Introduction}

The Boltzmann equation (BE) describes the space and time evolution of the velocity distribution function $f(\bm{x},\bm{\xi},t)$. The latter accounts for the number of fictitious particles at the location $\bm{x}$, time $t$, that are propagating with a given velocity $\bm{\xi}$:
% !!! : fictitious instead of fictive?
\begin{equation}
    \partial_t f + \xi_{\alpha}\partial_{\alpha}f = \Omega(f),
    \label{eq:BE}
\end{equation}
where Greek letters stand for space coordinates, and Einstein's summation rule is implied.
Roughly speaking, the BE (\ref{eq:BE}) illustrates the balance between transport and collision of particles. Assuming that $f$ is close to its equilibrium state $f^{eq}$, the collision term can be approximated by a relaxation mechanism
\begin{equation}
    \Omega(f) \approx -\dfrac{f-f^{eq}}{\tau},
    \label{eq:BGKcoll}
\end{equation}
which makes $f$ tend towards $f^{eq}$ under a relaxation time $\tau$, as originally proposed by Bhatnager, Gross and Krook, and more commonly known as BGK collision model~\cite{BHATNAGAR_PR_94_1954}. Even if the latter operator leads to a Prandtl number of unity and does not account for extra internal (rotational and vibrational) degrees of freedom, simple extensions exist to allow for the modeling of polyatomic gases with flexible Prandtl number~\cite{HOLWAY_PoF_9_1966,SHAKHOV_FD_3_1968,ANDRIES_EJMBF_19_2000,RYKOV_FD_10_1975}.

Since the 1950s, several types of deterministic solvers of the BGK-BE were proposed~\cite{MIEUSSENS_AIP_1628_2014}. For efficiency reasons, most of them rely on a (physical) discretization of the phase space, and can then be encompassed in the framework of discrete velocity methods (DVMs). The latter focus on the resolution of the discrete velocity Boltzmann equation (DVBE):    
\begin{equation}
    \partial_t f_i + \xi_{i\alpha}\partial_{\alpha}f_i = \Omega(f_i).
    \label{eq:DVBE}
\end{equation}
Contrary to the BE (\ref{eq:BE}), the DVBE (\ref{eq:DVBE}) is a set of partial differential equations of finite size $V$, where $V$ is the number of discrete velocities $\xi_i$ that compose the phase space of interest ($i \in \llbracket 1, V \rrbracket$).

\begin{figure*}[tbp!]
\centering
\includegraphics[width=.3\textwidth]{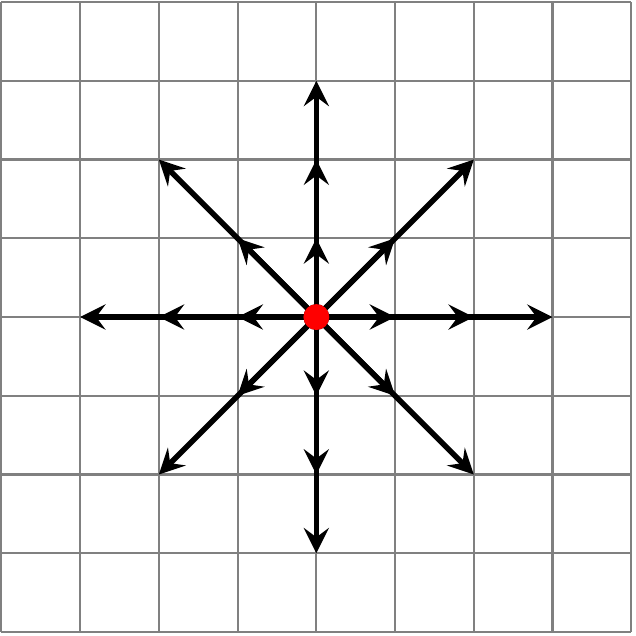}\quad
\includegraphics[width=.3\textwidth]{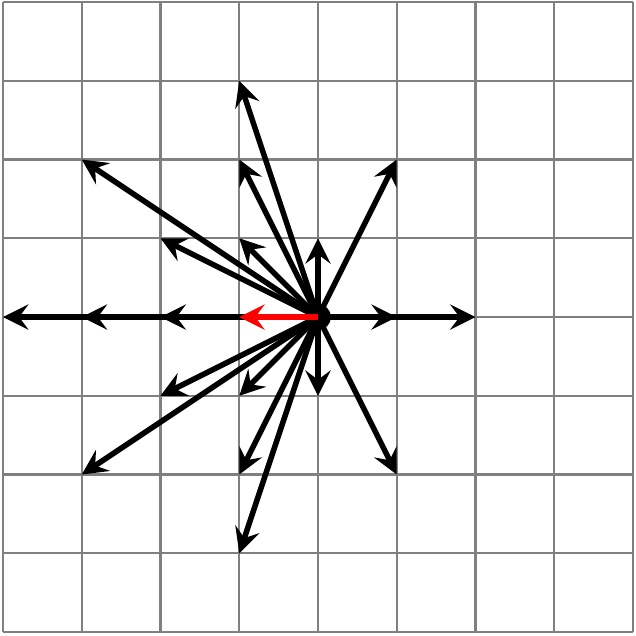}\quad
\includegraphics[width=.3\textwidth]{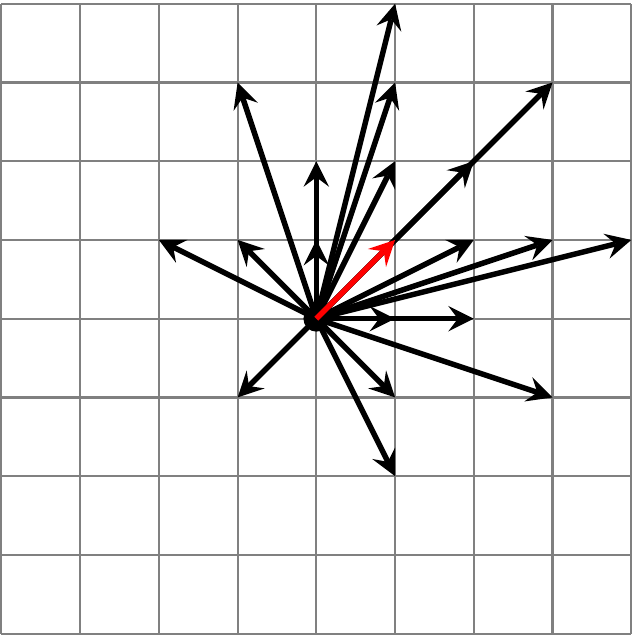}

\caption{Examples of on-grid shifted versions of the D2Q21 lattice: $c_{i\alpha}=\xi_{i\alpha}\sqrt{\theta} + U_{\alpha}$, with $U_{\alpha}\in \mathbb{Z}$ and $\theta=1$. The red arrow/dot corresponds to the reference population used for the shift: (0,0), (-1,0) and (1,1) from left to right.}
\label{fig:AdaptiveD2Q21}
\end{figure*}

For DVMs, the phase space discretization is chosen in such a way that all macroscopic velocity and temperature fluctuations of interest can be accounted for even in extreme conditions~\cite{MIEUSSENS_MMMAS_10_2000} --which usually leads to %large discrete velocity stencils. 
large velocity stencils compared to the standards of LBMs.
By further relying on Eulerian discretizations of the DVBE (finite-difference, finite-volume, etc), DVMs are able to simulate --in a stable manner-- flows with strong departures from equilibrium that are encountered, e.g., in rarefied conditions
~\cite{LETALLEC_TechReport_1997,CHARRIER_CRAS_326_1998,MIEUSSENS_MMMAS_10_2000,TITAREV_CICP_12_2012,BARANGER_JCP_257_2014,MIEUSSENS_AIP_1628_2014,ARISTOV_JCS_40_2020}. 
%Regarding LBMs, they also solve 
LBMs on the other hand also solve the DVBE (\ref{eq:DVBE}), but contrary to DVMs, they rely on the \textit{collide-and-stream} algorithm which is a very efficient Lagrangian-based numerical scheme~\cite{SCHORNBAUM_SIAM_38_2016}. %For the BGK-DVBE, it reads as
The corresponding BGK-DVBE reads as
\begin{equation}\label{eq:LBE-BGK}
    f_{i}\left(\boldsymbol{x}+\boldsymbol\xi_{i}, t + 1\right) = f_{i}\left(\boldsymbol{x}, t\right) \,-\, \dfrac{1}{\tau_f} \left(f_{i}-f_{i}^{eq}\right)\left(\boldsymbol{x}, t\right),
\end{equation}
where %LB units 
lattice units with unitary time and space step are implied, and $\tau_f = \tau + 1/2$.
In addition, the phase space discretization is chosen according to quadrature or moment-matching rules. This aims at recovering the physics of interest in terms of moments~\cite{SHAN_PRL_80_1998,SHAN_JFM_550_2006,GUO_Book_2013}, instead of velocity/temperature fluctuations. Hence, the size of the discrete phase space directly depends on the targeted physics. 
This is the reason why standard LBMs --that are based on 19 (or 27) discrete velocities-- are usually used as efficient alternatives to Navier-Stokes-Fourier (NSF) solvers for the simulation of isothermal weakly compressible flows~\cite{MANOHA_AIAA_2846_2015,BARAD_AIAA_4404_2017,RUMSEY_JA_56_2018,HOU_AIAA_2019_2555}, and also for more complex configurations~\cite{KRUGER_Book_2017,SUCCI_Book_2018,BAUER_CMA_CorrectedProof_2020,KRAUSE_CMA_CorrectedProof_2020,LATT_CMA_CorrectedProof_2020, DEROSIS_PoF_31_2019,DEROSIS_ARXIV_2020_01628}. 
%Eventually
To conclude the overview of methodologies, it is worth noting that discrete Boltzmann methods (DBMs) and discrete unified gas kinetic schemes (DUGKS), that share common features with both DVMs and LBMs, can also be used for the simulation of out-of-equilibrium phenomena that are encountered, e.g., with multi-component mixtures, reactive flows~\cite{LIN_CF_164_2016,LIN_PRE_99_2019, LIN_ARXIV_2020_02668} as well as multiscale rarefied gas flows
~\cite{GUO_PRE_88_2013,GUO_PRE_91_2015}. %As for 
Like DVMs, DBMs and DUGKS are based on Eulerian numerical schemes. Nonetheless, the moment-matching approach and quadrature rules are adopted (similarly to LBMs) for the derivation of the velocity stencil, which allows for a drastic reduction of its size.

All the aforementioned methodologies are usually based on static velocity discretizations. 
This means that if either strong out-of-equilibrium effects or important fluctuations of macroscopic quantities are encountered in a simulation, the accuracy and the stability of the numerical solver is at risk, because the velocity discretization does not match anymore the simulated physics. 
To tackle this issue, one can rely on adaptive velocity discretizations of the BE, which consist of sets of discrete velocities that self-adjust to local macroscopic quantities (such as velocity $\bm{u}$ and temperature $T$):
\begin{equation}\label{eq:discrete_norm_peculiar}
    c_{i\alpha} = \xi_{i\alpha} \sqrt{\theta} + u_{\alpha},
\end{equation}
with $\theta=T/T_0$ and $T_0$ being the reduced and lattice (or reference) temperatures, respectively. %$u_{\alpha}$ is the $\alpha$-component of the velocity vector $\bm{u}$.
This adaptive methodology was introduced in the 1990s to increase the accuracy and the stability of DVMs~\cite{NADIGA_JCP_112_1994,NADIGA_JCP_121_1995,NADIGA_JSP_81_1995,HUANG_IJMPC_08_1997}, as well as LBMs~\cite{SUN_PRE_58_1998,SUN_PRE_61_2000,SUN_JCP_161_2000,SUN_PRE_68_2003,SUN_CF_33_2004}, in the context of high-speed compressible flow simulations.
%The latter was recently revived through improved space interpolation and reconstruction strategy of missing information~\cite{DORSCHNER_PRL_121_2018,ZAKIROV_KIAM_Preprint_Eng_2019,ZIPUNOVA_DSFD_2020}.
Interestingly, adaptive DVMs are \textit{interpolation-free} solvers for the evolution of discrete populations due to their Eulerian nature. 
On the contrary, adaptive LBMs rely on a Lagrangian-based scheme (\textit{collide-and-stream}) meaning that populations are not necessarily streamed from one grid node to another one, e.g., if the discrete velocity components are non-integer. Hence, the algorithm must be supplemented with a space interpolation step to recover populations at grid nodes. This additional step is computationally expensive, potentially prone to anisotropy issues (depending on the considered stencil of interpolation points), and deteriorates the natural parallel efficiency of LBMs. This reduces the interest one might have in moving from NSF solvers to adaptive LBMs for the simulation of high-speed compressible flows.%, especially in an industrial context where the efficiency of the numerical solver is of paramount importance. 

To get rid of the interpolation step, one must ensure that %the on-grid condition is verified whatever the flow conditions ($c_{i\alpha}\in\mathbb{Z}$). This can be achieved through jumps between lattices whose discrete velocities component are integer valued, which guarantees an \textit{exact} streaming step.
all discrete velocity components are integer valued in any flow condition ($\xi_{i\alpha},c_{i\alpha}\in\mathbb{Z}$), which guarantees an \textit{on-grid} streaming step.
These conditions on the velocity/temperature shifts of discrete velocities are illustrated in Fig.~\ref{fig:AdaptiveD2Q21} for the D2Q21 lattice --whose characteristics can be found in the work by Zhang et al.~\cite{ZHANG_PRE_74_2006}. Yet, jumping from a lattice to another one in a discrete manner imposes new restrictions and challenges on the discrete stencils %so that the transition can be done 
to guarantee that the transition is achieved in an accurate and stable manner. 
Such an investigation is the starting point of our work, which is organized as follows. In Section~\ref{sec:choice_of_EDF}, the viability of the transition between velocity discretizations is investigated by looking at operability range overlaps for LBMs based on polynomial and numerical equilibria. After finding a good compromise between stability, accuracy and efficiency, shifting criteria and interpolation-free strategies are discussed in Section~\ref{sec:conservative_strategies}. The ability of interpolation-free adaptive LBMs to handle both linear (propagation of shear, entropy and acoustic waves) and non-linear regimes (1D and 2D Riemann problems) is then investigated in the low-viscosity limit (Section~\ref{sec:numerical_applications}). Conclusions are drawn % on the present proof-of-concept
in Section~\ref{sec:conclusion}. Eventually, a number of appendices are provided to help the reader further understanding concepts used in this work: \rcol{(\ref{app:maxwell_hermite}) approximation of the Maxwellian in a given reference frame}, (\ref{app:VN}) linear stability analyses, (\ref{app:numericalEq}) detailed description of numerical equilibria, \rcol{(\ref{app:errorMacro}) preliminary study regarding the choice of lattice and set of constraints}, and (\ref{app:PrandtlExtension}) extension to variable Prandtl number. %Chapman-Enskog expansion of reconstructed populations.% at the transition interface.

\section{\label{sec:choice_of_EDF} Realizability of phase space transitions}

\subsection{\label{subsec:motivation}Motivation}

In the LB framework, the velocity space discretization is most commonly based on stencils that are static (in space and time), symmetric, and consequently, centered around a population at rest. 
This implies that these lattices are dedicated to the simulation of  quasi-quiescent flows with little temperature fluctuations, i.e., $(\bm{u},T)\approx (\bm{0},T_0)$~\cite{MIEUSSENS_MMMAS_10_2000}. 
This is further supported by the fact that corresponding equilibrium distribution functions (EDFs) are usually based on \rcol{polynomial} expansions of the Maxwellian about $(\bm{u},T)= (\bm{0},T_0)$~\cite{GUO_Book_2013}. When deviating from the reference state $(\bm{0},T_0)$, both the accuracy and the stability of these LBMs are impacted, and this can be quantified through macroscopic error evaluations and linear stability analyses~\cite{HOSSEINI_PRE_100_2019}. \rcol{In fact, one can already identify for a continuous phase space issues induced by: (1) the approximation of the Maxwellian and (2) the associated reference frame. The interested reader may refer to Appendix~\ref{app:maxwell_hermite} for an in-depth discussion regarding the latter point.} 
% !!! Would you like me to give you plots of the MB distribution and its polynomial approximations for different velocities, with and without shift to clearly show that with the adaptive discretization one can match the MB distribution even with low order quadratures?

To make sure the correct physics can be simulated in a stable and accurate manner, \rcol{one should} adapt the direction and norm of discrete velocities to the mean flow conditions~\cite{MIEUSSENS_MMMAS_10_2000}, as shown in Fig.~\ref{fig:AdaptiveD2Q21}% -- without increasing their number in order not to increase the computational cost of the solver
. In addition, one must ensure that the transition between two states of the same lattice can be done in a stable and accurate manner. Henceforth, corresponding requirements for stable and accurate phase space transitions are investigated. They directly depend on how the LBM is derived, and more precisely, on the type of EDF used to recover the macroscopic behavior of interest. For LBMs based on polynomial EDFs, it is proposed to perform linear stability analyses to find the minimal lattice size allowing for stability domain overlaps between two shifted versions of the same lattice. Regarding numerical EDFs, the latter methodology cannot be used anymore, and instead, the convergence of the root-finding solver is considered as an alternative criterion for the evaluation of stability domain overlaps.

\subsection{\label{subsec:polynomial_EDF}Realization based on polynomial discrete equilibria}

A variety of systematic approaches to approximate the Maxwell-Boltzmann equilibrium in a discretized phase space have been developed. Of these methods, two have become rather popular~\cite{SHAN_PRL_80_1998,SHAN_JFM_550_2006,KARLIN_PA_389_2010,GUO_Book_2013}: (a) truncated Hermite expansion and (b) moment-matching. While they both converge to the same discrete equilibrium state for tensor-product-based stencils (e.g. DdQ$3^{\mathrm{d}}$), the latter is more flexible %, straight-forward
and allows to derive a discrete equilibrium for almost any stencil structure. As clearly indicated by its name, it mainly consists in matching the moments of the discrete equilibrium state to those of the Maxwell-Boltzmann distribution up to the highest order of interest. 
 
\subsubsection{\label{subsubsec:polynomial_EDF_construction}Polynomial EDF construction and corresponding lattices}

For the scheme to correctly recover the targeted physics, the number of discrete (shifted) velocities in the stencil $c_i$ must be higher than the minimum number of constraints, {i.e.}, moments of the EDF needed to match their continuous counterparts. As such the discrete equilibrium construction process comes down to solving a system of algebraic equations of the following form:
\begin{equation}\label{eq:polyEqSys}
    \bm{G f^{eq}} = \bm{M}^{\mathrm{MB}},
\end{equation}
where $\bm{M}^{\mathrm{MB}}$ is the constraints vector containing the continuous moments of the Maxwellian, with the ${n}^{\mathrm{th}}$ moment ($n=p+q\in \mathbb{N}$) defined as:
\begin{equation}
    M^{\mathrm{MB}}_{pq} = \int c_{x}^{p} c_{y}^{q} f^{\mathrm{MB}} d\bm{\xi}.
\end{equation}
and
\begin{equation}
    f^{\mathrm{MB}}=\dfrac{\rho}{(2\pi \theta)^{d/2}}\exp\bigg[-\dfrac{(\xi-u)^2}{2\theta}\bigg],
\end{equation}
where $d$ is the number of physical dimensions. Assuming $\bm{G}$ is invertible, the system of equations~(\ref{eq:polyEqSys}) can be solved, and the discrete (shifted) equilibria are then obtained as:
\begin{equation}
    \bm{f^{eq}} = \bm{G}^{-1}\bm{M}^{\mathrm{MB}}.
\end{equation}
For a one-dimensional physical space, the DVBE correctly recovers the NSF dynamics if and only if the first five moments of the EDF are exactly matched. If we restrict the study to tensorial products of 1D velocity stencils, then the smallest stencils that satisfies these conditions are of the form DdQ$5^{\mathrm{d}}$. A well-known illustration of this moment-matching requirement is the third-order family of stencils (DdQ$3^{\mathrm{d}}$) widely used in the LB community for low-Mach isothermal flows. The latter restriction results from the number of degrees of freedom in the system that prevents these lattices to correctly recover moments higher than two.

\subsubsection{\label{subsubsec:polynomial_EDF_operability_range}Assessment of operability range: linear stability domain}

%\bcol{Check this section and add references to previous works and corresponding appendix. Say it is sufficient to run 1D LSA because in 2D and 3D more waves are present and lead to more instabilities~\cite{RENARD_ARXIV_2020_08477}}

\begin{figure*}[btp!]
	\centering
	%\hspace{-0.1\textwidth}
	%\begin{subfigure}{0.3\textwidth}
	%	 \includegraphics[trim=22 22 340 210, clip]{./figures/min_stable_order.eps}
	%\end{subfigure}
	\includegraphics[width=.75\textwidth]{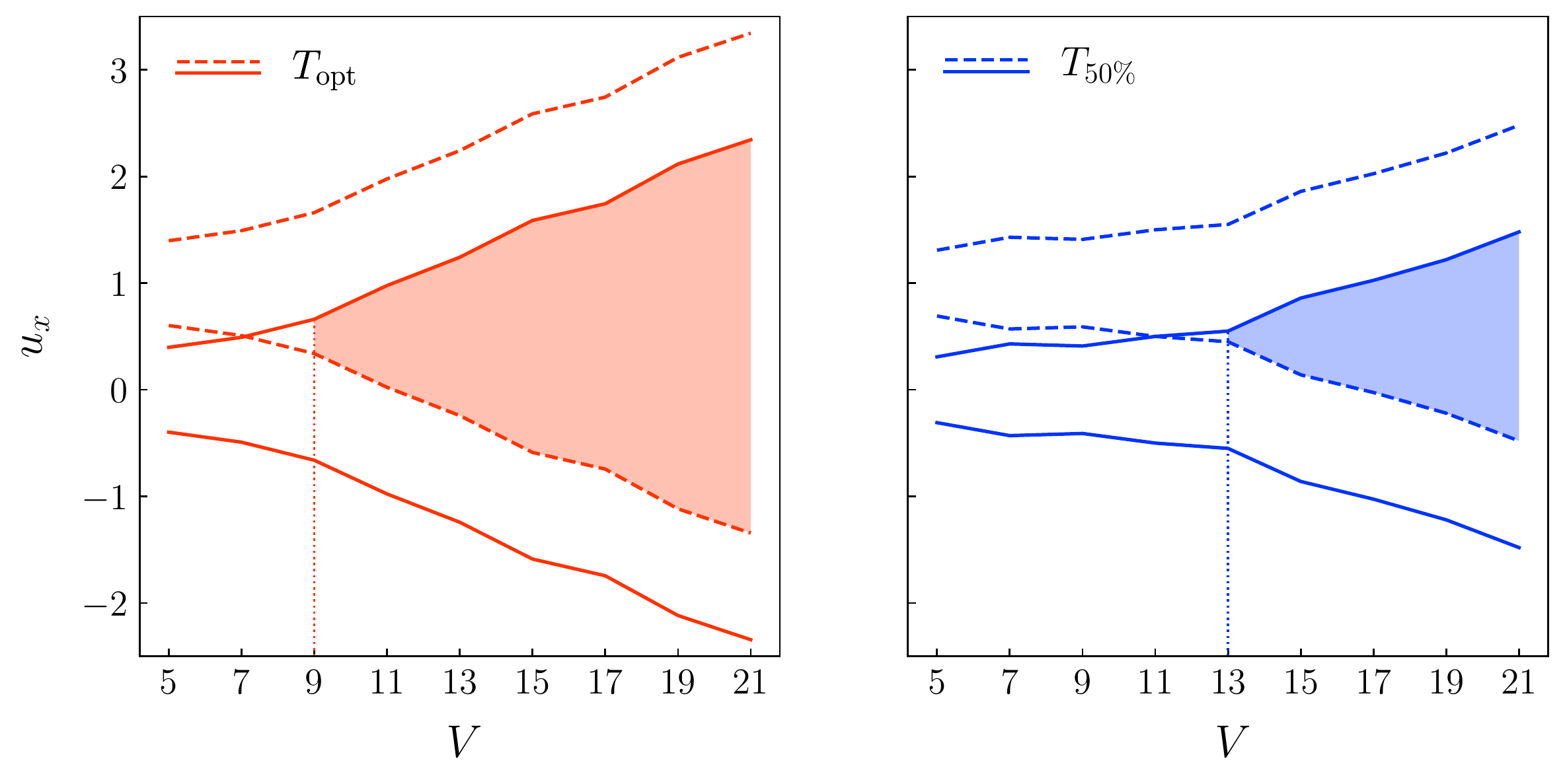}
\caption{Minimal stability condition which allows adaptive shifting with LBMs based on polynomial equilibria. Stability domain obtained for different 1D lattices of size $V$: (solid) $U_x=0$, and (dashed) $U_x=1$. Configurations $T_{\mathrm{opt}}$ and $T_{50\%}$ corresponds to the maximum achievable velocities at optimal reference temperature, and with at least 50 percent variation in temperature, respectively. The dotted lines represent the minimal stencil required for an overlap of $\Delta u_x \geq 0.1$ (shaded region). This leads to the D1Q9 and D1Q13 lattices for $T_{\mathrm{opt}}$ and $T_{50\%}$ respectively.}
\label{Fig:min_stable_order}
\end{figure*}

The operability range of a given DVBE solver, targeting the Navier-Stokes-Fourier (NSF) system of equations, can be assessed following the Lax-Richtmyer equivalence theorem conditions, namely stability and consistency.

The latter can be readily evaluated by looking at the deviations of moments (appearing at the NSF level) of the discretized equilibrium function from their continuous counter-parts. This tool can be especially useful for low order stencils that fail to impose all constraints tied to the NSF level moments. 
However, for larger stencils that explicitly enforce physical constraints on all moments intervening at the NSF level, deviations reduce down to zero and are therefore useless in assessing the operation range of the solver. 
While a necessary condition for the equivalence, consistency is not sufficient and must be completed by a proof of stability of the numerical scheme. 

The linear stability analysis (also called von Neumann or Fourier analysis) is a rather popular and widely accepted approach to assess the stability and accuracy of a given linear system of discrete equations~\cite{VONNEUMANN_Book_1966,WOLF_GLADROW_Book_2004,HIRSCH_Book_2007}. In the LB context, this tool was used for various comparative studies, and to quantify the impact of several parameters: numerical discretization of the DVBE~\cite{WILDE_IJNMF_90_2019}, spatial filtering~\cite{RICOT_JCP_228_2009}, accuracy and efficiency comparison with NSF solvers~\cite{MARIE_JCP_228_2009}, optimization of multi-relaxation time collision models~\cite{LALLEMAND_PRE_61_2000,XU_JCP_230_2011,XU_JCP_231_2012,CHAVEZMODENA_CF_172_2018,HOSSEINI_PRE_99_2019b}, stabilization mechanism of collision models~\cite{DUBOIS_CRM_343_2015,COREIXAS_RSTA_378_2020,WISSOCQ_ARXIV_2020_07353}, shifted stencils~\cite{HOSSEINI_PRE_100_2019}, etc.

While intended for linear equations, it is also widely used to assess the stability properties of non-linear systems, by replacing the target equations with a linearized version~\cite{VONNEUMANN_Book_1966,WOLF_GLADROW_Book_2004,HIRSCH_Book_2007}. For LBMs, this boils down to the linearization of the collision term since the convective part is already expressed in a linear form. 
While -strictly speaking- the linearization limits the validity of the analysis to the linear regime, one can argue that in practice this just changes the value of the outcome from \emph{sufficient} to merely a \emph{necessary} condition. 
Based on the latter assertion, the linear stability analysis is still relevant as it establishes upper bounds for the operability range of the solver, making it especially interesting for higher-order stencils with polynomial equilibria.% where the analysis based on deviation of the EDF moments becomes somewhat obsolete.

%As such, to justify the latter choice of numerical equilibria, we establish a trend for the linear stability domain as a function of the order of the stencil in the context of polynomial equilibria.
As such, we establish a trend for the linear stability domain as a function of the order of the stencil in the context of polynomial equilibria.
Following the approach described in~\cite{HOSSEINI_RSTA_378_2020} and briefly recalled in Appendix~\ref{app:VN}, the analysis is performed in the $\{u, \theta, \tau\}$ parametric space for 1D symmetric stencils of different orders, both with and without shifts. The largest velocity resulting in a stable system in the limit of vanishing viscosity is reported as the stability domain of the stencil. 
However it is common knowledge that while allowing for larger Mach numbers, larger stencils result in narrower temperature ranges.
As such a second stability domain, based on the maximum velocity allowing for fifty percent variation in temperature is also reported. The obtained results for shifted and non-shifted stencils are shown in Fig.~\ref{Fig:min_stable_order}.
It can be readily observed that the smallest stencil even allowing for a stable unit shift is the D1Q9 lattice ($\xi_{i\alpha} \in \{0,\pm 1,\pm 2, \pm 3, \pm 4\}$). For the stencil to allow for a stable shift and -more or less- pronounced temperature variations the stencil size goes up to $\xi_{i\alpha} \in \{0,\pm 1,\pm 2, \pm 3, \pm 4, \pm 5, \pm 6\}$, making it rather inefficient in terms of memory consumption, processing power and communication overhead in parallel computations.

Therefore, while our study of shifted stencils with polynomial equilibria remains of theoretical interest, practical application are better served with the approach of numerical equilibria described below.

\subsection{\label{subsec:numerical_EDF}Realization based on numerical discrete equilibria}

While polynomial EDFs impose strong constraints on the velocity discretization through the moment-matching approach, numerical EDFs allow for the derivation of quadrature-free LBMs, hence providing more freedom regarding the size of the velocity stencil. During the past three decades, the latter EDFs have been extensively used in the context of rarefied gas flow simulations for both static and adaptive phase space discretizations~\cite{LETALLEC_TechReport_1997,CHARRIER_CRAS_326_1998,DUBROCA_CRAS_329_1999,MIEUSSENS_JCP_162_2000,MIEUSSENS_MMMAS_10_2000,MIEUSSENS_PoF_16_2004,BARANGER_JCP_257_2014,MIEUSSENS_AIP_1628_2014}. Hereafter, they are introduced as interesting alternatives to polynomial EDFs for the transition between two velocity space discretizations of smaller size.

\subsubsection{\label{subsubsec:numerical_EDF_construction}Construction of quadrature free discrete equilibria}

This kind of EDF results from the minimization of the $H$-functional 
 ~\cite{OTTINGER_RSTA_378_2020}
 \begin{equation}\label{eq:Hfunction}
 H=\sum_i f_i [\ln{(f_i/a})]
 \end{equation}
under the constraints
\begin{equation}
G_{pq}=\sum_i f_i^{eq} \xi_{ix}^p\xi_{iy}^q - M_{pq}^{\mathrm{MB}}=0\label{eq:constraints}.
\end{equation}
and reads as
\begin{equation}\label{eq:exponential}
 f_i^{eq} = a \exp[-(1+\textstyle{\sum_{p,q}}\lambda_{M_{pq}^{\mathrm{MB}}}\xi_{ix}^p\xi_{iy}^q)],
\end{equation}
where $\lambda_{M_{pq}^{\mathrm{MB}}}$ are the Lagrange multipliers corresponding to the constraints~(\ref{eq:constraints}). Following our previous work~\cite{LATT_RSTA_378_2020}, the prefactor $a=\rho$ is adopted hereafter. 

\begin{figure*}[btp!]
	\centering
\includegraphics[width=.85\textwidth]{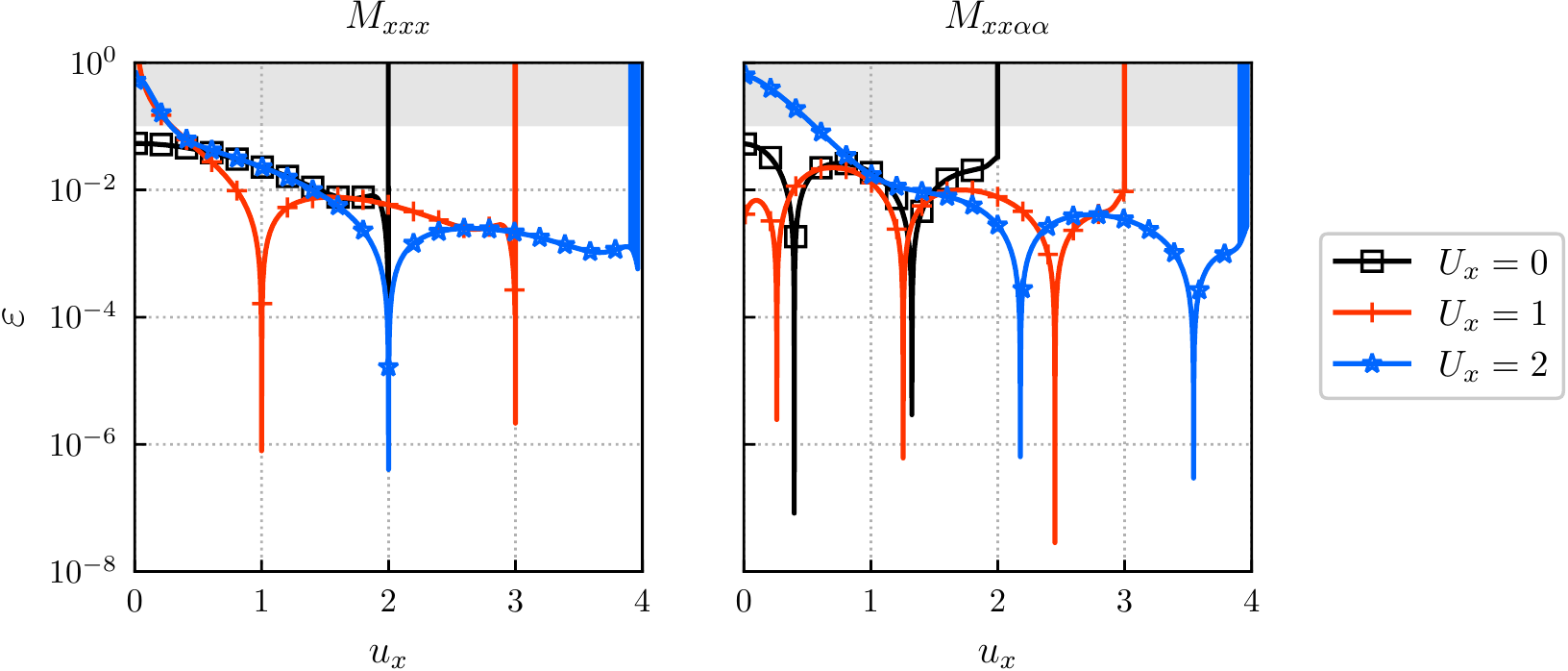}
\caption{Velocity and shift impact on macroscopic errors $\epsilon$ for the D2Q21 lattice with 8-moment approach and $T = 0.7$.
The flow propagates along the x-axis, and the grey zone starts at $\epsilon = 10 \%$. The velocity shift induces a shift of the upper limit of the root-finding solver convergence in terms of velocity. This eventually leads to large stability domain overlaps between different shifted versions of the D2Q21 lattice.}
\label{Fig:numEDF_stability}
\end{figure*}

%The numerical equilibrium (\ref{eq:exponential}) is less ``user-friendly'' than its polynomial counterpart, in the sense that its related macroscopic behavior cannot be derived in an exact manner anymore. 
%As the numerical equilibrium~(\ref{eq:exponential}) cannot impose an exact value for unconstrained equilibrium moments, the accuracy of physical phenomena related to these moments cannot be assessed theoretically, and we need to conduct numerical tests to determine their accuracy limits.
Assuming the set of constraints (\ref{eq:constraints}) corresponds to the conservation of mass, momentum and total energy (the zeroth, first and trace of second-order Maxwellian moments), the Chapman-Enskog (CE) expansion gives~\cite{CHAPMAN_Book_3rd_1970}:
\begin{equation}
\begin{array}{c}
\partial_t \rho + \partial_{\chi}(\rho u_{\chi})=0,\\[0.1cm]
\partial_t (\rho u_{\alpha}) + \partial_{\beta}(\rho u_{\alpha}u_{\beta} + p \delta_{\alpha\beta})+\Delta_2 =\partial_{\beta}(\Pi_{\alpha\beta}) + \Delta_3 ,\\[0.1cm]
\partial_t (\rho E) + \partial_{\alpha}((\rho E +p)u_{\alpha}) + \Delta_3^{tr}=\partial_{\alpha}(\Phi_{\alpha}) + \Delta_4^{tr}.
\end{array}
\label{eq:NSF_13moments}
\end{equation}
$\Delta_n$ and $\Delta_n^{tr}$ are errors that emerge because their corresponding constraints (moments of order $n$ or their trace) are not accounted for in the computation of the exponential equilibrium (\ref{eq:exponential}). While diffusive errors ($\Delta_3$ and $\Delta_4^{tr}$) can usually be neglected at moderate or high Reynolds numbers~\cite{LATT_RSTA_378_2020}, this is not the case for deviations related to convective terms unless one adopts large velocity discretizations. %\rcol{(see Sec. where Q49 is not able to recover the correct thermal diffusion for $u>0.5$ when relying on 4 moments only)}. 
In order to derive efficient and accurate LBMs based on the numerical equilibrium (\ref{eq:exponential}), one of the following sets of constraints should be adopted~\cite{LEVERMORE_JSP_83_1996,OTTINGER_RSTA_378_2020,LATT_RSTA_378_2020}:
\begin{itemize}
    \item \underline{8-moment}:
    \begin{equation}\label{eq:8moments}
    (1,\:\xi_{i\alpha},\:\xi_{i\alpha}\xi_{i\beta},\:\xi_{i\chi}\xi_{i\chi}\xi_{i\alpha})    
    \end{equation}    
    \item \underline{9-moment}:
    \begin{equation}\label{eq:9moments}
    (1,\:\xi_{i\alpha},\:\xi_{i\alpha}\xi_{i\beta},\:\xi_{i\chi}\xi_{i\chi}\xi_{i\alpha},\:\xi_{i\chi}\xi_{i\chi}\xi_{i\eta}\xi_{i\eta})    
    \end{equation}
    \item \underline{10-moment}:
    \begin{equation}\label{eq10moments}
    (1,\:\xi_{i\alpha},\:\xi_{i\alpha}\xi_{i\beta},\:\xi_{i\alpha}\xi_{i\beta}\xi_{i\gamma})    
    \end{equation}
    \item \underline{11-moment}:
    \begin{equation}\label{eq:11moments}
    (1,\:\xi_{i\alpha},\:\xi_{i\alpha}\xi_{i\beta},\:\xi_{i\alpha}\xi_{i\beta}\xi_{i\gamma},\:\xi_{i\chi}\xi_{i\chi}\xi_{i\eta}\xi_{i\eta})    
    \end{equation}  
    \item \underline{13-moment}:
    \begin{equation}\label{eq:13moments}
    (1,\:\xi_{i\alpha},\:\xi_{i\alpha}\xi_{i\beta},\:\xi_{i\alpha}\xi_{i\beta}\xi_{i\gamma},\:\xi_{i\chi}\xi_{i\chi}\xi_{i\alpha}\xi_{i\beta})    
    \end{equation} 
\end{itemize}
%where Greek subscripts stand for 2D coordinates $x$ or $y$, and the repeated-index summation convention is implied.
The set of eight constraints (\ref{eq:13moments}) is the minimal configuration allowing to \textit{accurately} simulate convective phenomena ($\Delta_2=\Delta_3^{tr}=0$). Further accounting for constraints related to $M_{\alpha\beta\gamma}^{\mathrm{MB}}$ and $M_{\alpha\beta\chi\chi}^{\mathrm{MB}}$, one ends up with the 13-moment approach~(\ref{eq:13moments}) which \textit{exactly} leads to the Navier-Stokes-Fourier level of physics in the continuum regime.

\subsubsection{\label{subsubsec:numerical_EDF_operability_range}Operability range based on the convergence of the root-finding algorithm and non-linear kinetic stabilization}

Increasing the number of constraints automatically leads to better macroscopic properties. 
Nevertheless, this generally comes at the cost of a reduced operating range in terms of velocity and temperature fluctuations. 
This is explained by the fact that increasing the number of constraints puts more effort on the root-finding algorithm used for the computation of the numerical equilibrium. Depending on the type of algorithm considered (bisection, secant, Newton, etc), the robustness of the resulting LBM can be strongly impacted. 
Algorithms based on exact formulations of the Jacobian matrix lead to wider stability ranges, and we noticed that the larger the lattice size, the wider the stability domain of the LBM. 
%This is because increasing the number of discrete velocities gives more flexibility for the convergence of the root-finding solver. 
Yet, a trade-off has to be made between lattice size and number of constraints in order to obtain an accurate, robust and efficient LBM. 

In the early stage of this work, several lattices were studied (D2Q9, D2Q13, D2Q17, D2Q21, D2Q49, D2Q81) in combination with a large number of constraints, i.e., $M\in\{4,8,9,10,11,13\}$. In the context of supersonic flow simulations, and following the methodology proposed in our previous work~\cite{LATT_RSTA_378_2020}, a good trade-off between accuracy, stability and efficiency was found with the D2Q21 lattice~\cite{ZHANG_PRE_74_2006} based on the EDF computed via $M=8$ constraints (\ref{eq:8moments}) \rcol{-- the interested reader may refer to Appendix~\ref{app:errorMacro} for a brief summary of this preliminary study}. In particular, \rcol{this set of constraints} ensures the correct macroscopic behavior related to convective phenomena. Regarding diffusive errors, they \rcol{should remain as much as possible under} $\epsilon = 10 \%$, where the deviation from macroscopic moments is computed through
\begin{equation}
    \epsilon = \dfrac{\big\vert M_{pq}^{\mathrm{MB}} - M_{pq}^{eq}\big\vert}{M_{pq}^{\mathrm{MB}}}, 
\end{equation}
as originally proposed by Kornreich and Scalo~\cite{KORNREICH_PD_69_1993}. To illustrate this point, deviations obtained with the present LBM (D2Q21, $M=8$ and reference temperature $T_0=0.7$) are compiled in Fig.~\ref{Fig:numEDF_stability} for two moments related to diffusive phenomena, namely, the third-order moment
\begin{equation}
    M_{xxx}^{\mathrm{MB}} = M_{30}^{\mathrm{MB}} = \rho u_x (u_x^2 + 3 T),
\end{equation} 
and the trace of the fourth-order moment \begin{equation}
    M_{xx\alpha\alpha}^{\mathrm{MB}} = M_{40}^{\mathrm{MB}} + M_{22}^{\mathrm{MB}} = 2\rho[( E + 2  T) u_x^2 +  T (E + T)].
\end{equation} 
 One can observe that error levels remain under the $10\%$ threshold for a large range of Mach numbers. In addition, by adapting the lattice in a discrete manner, it is confirmed that the root-finding solver can converge for even higher values of the Mach number while leading to acceptable deviations with respect to the targeted physics. More importantly, large overlapping zones are now present for several shifts of the D2Q21 lattice which would solve the problem encountered with polynomial equilibria, i.e., the need for large lattices to allow the transition between reference states.

Assuming that non-equilibrium phenomena remain negligible, the convergence of the root-finding solver can be considered as a necessary and (almost) sufficient condition for the stability of the present approach in either its static- or adaptive-state form. In this work, an improved version of the Newton-Raphson algorithm was used to further increase the convergence domain of the root-finding solver. It is based on Powell's Hybrid method -- available in the GSL library~\cite{GSL2020} -- that combines the fast convergence of Newton's approach with a gradient descent (see Chaps 6 and 7 in~\cite{RABINOWITZ_Book_1970} for more details regarding this algorithm). Eventually, to further improve the stability in case of strong departure from equilibrium and/or due to under-resolved conditions, we propose to also include a non-linear stabilization technique that shares some similarities with flux limiters~\cite{LATT_RSTA_378_2020}. Its main characteristics are briefly recalled in Section~\ref{subsubsec:stabilization}.

\section{\label{sec:conservative_strategies}Coupling strategies for shifted LBMs}

%\subsection{\label{subsec:motivation}Motivation}

In the previous section, the possibility to switch between different versions of the same lattice in a stable manner has been detailed. We now focus our discussion on the two main technicalities related to the coupling of different LB solvers: (1) when %should we 
to switch between the two solvers, and (2) how %will the information be exchanged 
to exchange the information at the interface between them. The former will first be tackled by discussing local criteria for the shift of velocity space discretization. The second issue will be dealt with investigating the different strategies for the reconstruction of missing populations at the interface between two states of the (same) velocity discretization. 
%All of this is discussed 
This discussion is conducted in the context of polyatomic gas flow simulations in which rotational (and vibrational) degrees of freedom are accounted for through a second set of populations $g_i$~\cite{RYKOV_FD_10_1975,DUBROCA_ESAIM_10_2001,TITAREV_ECCOMAS_2006,NIE_PRE_77_2008,FRAPOLLI_PhD_2017,LATT_RSTA_378_2020}, as discussed in Section~\ref{subsubsec:underlying_LB_scheme}.

\subsection{\label{subsec:shifting_criterion}Criteria for lattice shifting}

Hereafter, we focus on criteria that can be used to decide which version of the lattice better fits the physics of interest at a given node $\bm{x}$ and time $t$. 

\subsubsection{\label{subsubsec:breakdown_criterion}Breakdown criteria}

In the literature dedicated to the coupling of LBMs with other kinetic based solvers, e.g., for rarefied gas flow simulations (DSMC~\mbox{\cite{DISTASO_JCS_17_2016,DISTASO_JCS_17_2016,DISTASO_CF_172_2018} }, DVMs~\mbox{\cite{ARISTOV_EUCASS_2017,ARISTOV_AIPCP_2132_2019,ARISTOV_JCS_40_2020} } or high-order LBMs~\mbox{\cite{MENG_PRE_83_2011} }), valuable information is found % In the literature dedicated to the coupling of LBMs with other kinetic based solvers, e.g., for rarefied gas flow simulations (DSMC~\cite{DISTASO_JCS_17_2016,DISTASO_JCS_17_2016,DISTASO_CF_172_2018}, DVMs~\cite{ARISTOV_EUCASS_2017,ARISTOV_AIPCP_2132_2019,ARISTOV_JCS_40_2020} or high-order LBMs~\cite{MENG_PRE_83_2011}), valuable information is found 
regarding static and dynamic domain decomposition relying on particular criteria such as the breakdown of the continuum assumption. 

In the static case, more advanced solvers are used close to walls in order to capture out-of-equilibrium effects, such as velocity and temperature jumps, which cannot be simulated by continuum based solvers (NSF or low-order LBMs), at least, for Knudsen numbers close to unity. The domain decomposition is then %built in a manual manner, as done, e.g., for grid mesh refinement
designed manually, \rcol{as similarly} done for mesh refinement.~\cite{DUPUIS_PRE_67_2003,LAGRAVA_JCP_231_2012,GENDRE_PRE_96_2017,ASTOUL_ARXIV_2020_14887}. The latter approach assumes %that one knows in advance which parts of the simulation domain will be related to out-of-equilibrium phenomena.
an a priori knowledge of domain portions related to out-of-equilibrium phenomena. %In the most general case, such an information is not available, and a criterion is then required to determine the domain decomposition in a dynamic manner.
When this information is unavailable, a dynamic domain decomposition criterion is necessary. Most breakdown criteria are based on the (local) Knudsen number, gradients of macroscopic quantities, or deviations with respect to (non-)equilibrium moments/populations~\cite{BOYD_PoF_7_1995,TIWARI_JCP_144_1998,WANG_PoF_15_2003,LOCKERBY_RSTA_465_2009,MENG_PRE_89_2014,BARANGER_JCP_257_2014}. 

%Even if the above-mentioned works are dedicated to rarefied gas flow simulations, they proved that rather simple criteria are sufficient to build the domain decomposition in either a static or dynamic manner. Regarding the criteria themselves, they cannot be used as they are in the present context because they are not based on the same philosophy. Here, we want to move from a lattice to one of its shifted versions in order to adjust references quantities, such as velocity and or temperature, which will improve the stability and accuracy of the resulting solver. To achieve such a goal, it then seems better to rely on criteria that are based on local values of macroscopic quantities of interest.

A criterion for the local shift of velocity stencils, as targeted in this paper, seems however more straightforward to formulate, as it is sufficient to formulate bounds on the macroscopic quantities of interest.

\subsubsection{\label{subsubsec:shifting_criterion}Criteria based on macroscopic quantities}

In the context of numerical EDFs, it is possible to evaluate at the same time deviations from the macroscopic behavior of interest, as well as stability limits, through studies similar to those conducted in Fig.~\ref{Fig:numEDF_stability}. This is the starting point for our search for adequate phase space transition criteria.

The first criterion one might think of is a dimensionless one, namely, the Mach number $\mathrm{Ma}=u/\sqrt{\gamma_r T}$, with $u$ the norm of the velocity vector $\bm{u}$, and $\gamma_r$ the specific heat ratio. 
%Even if this criterion is pretty convenient because it has the same definition for both physical and LB unit systems, it is rather deceiving in the sense that it hides the true nature of instabilities.
Although this criterion is rather convenient, as it is based on a dimensionless quantity, it fails to reveal the true nature of instabilities.
%Taking the example of the D2Q21-LBM based on the numerical equilibrium with $M=8$, the root-finding solver becomes unstable for an uniform flow propagating along the $x$-axis 
For example, a simulation based on the D2Q21-LBM with numerical equilibirum (M=8) shows that numerical stability is limited by a range of lattice-unit velocities rather than the Mach number. Indeed, assuming a uniform flow propagating along the x-axis at $\mathrm{Ma} = 1.85$ with $T=0.7$ ($\theta=1$) and $\gamma_r = 5/3$, the root-finding solver becomes unstable for $u_x\geq2$. By changing the value of $T$ to $0.6$ ($\theta=1$), one obtains a higher stability limit $\mathrm{Ma} = 2$ whereas the threshold remains the same for the velocity. 
Hence, the macroscopic velocity is in fact more appropriate than the Mach number to decide if the lattice must be locally adjusted.

%By repeating the above experiment, one might think that it is possible to arbitrarily increase the Mach number upper limit by simply decreasing the temperature. 
The above experiment might lead to the conclusion that the Mach number upper limit can be increased arbitrarily by decreasing the reference temperature of a simulation.
%Yet, this assumes the root-finding solver converges whatever the value of $T$, which is not the case in practice. 
In practice, however, the temperature $T$ is limited by the stability range of the root-finding solver.
%Hence, the macroscopic temperature $T$ can be considered as a possible criterion to locally adjust the lattice. 
This observation could lead to the desire to adjust the reference temperature from a portion of the domain to another, just as it is done for the reference velocity of the lattice. 
However, temperature fluctuations remain sufficiently low in the conducted experiments ($T_{\max}/T_{\min}< 2$) so that it is not necessary to adjust the reference temperature through a phase space transition.

In the end, the criterion used to move from a lattice to one of its shifted version is
\begin{equation}\label{eq:criterion}
    n - 0.5 < u_{\alpha} \leq n + 0.5 \quad \Longleftrightarrow \quad U_{\alpha} = n
\end{equation}
where $n\in \mathbb{Z}$. For the D2Q21-LBM based on the numerical EDF with $M=8$, the above intervals ensure that the previous and the updated lattices will share similar macroscopic deviations. This is justified, once again, through analyses that are conducted in Fig.~\ref{Fig:numEDF_stability} where error levels are almost identical for $U_x=0$ and $1$ close to $u_x=0.5$ (the same goes for $U_x=1$ and $2$ close to $u_x=1.5$). Taking the example of lattices considered in Fig.~\ref{fig:AdaptiveD2Q21}, we then end up with the lattice shifts $(U_x,U_y)=(0,0)$, $(-1,0)$ and $(1,1)$ for the following macroscopic conditions $(-0.5,-0.5)<(u_x,u_y)\leq (0.5,0.5)$, $(-1.5,-0.5)<(u_x,u_y)\leq (-0.5,0.5)$ and $(0.5,0.5)<(u_x,u_y)\leq (1.5,1.5)$, respectively.
%Eventually, to avoid any hysteresis effect that might appear for half-integer mean flow velocities (as it will be the case for several benchmark tests in Section~\ref{subsec:preliminary_investigations}), we will fixed $n_0=0.51$. 

Eventually, to reduce the number of interface cells, or the frequency of change of
reference velocity, it can be useful to use velocity ranges of a width
larger than 1 (e.g., a range $[-1, 1]$ for \rcol{$U_{\alpha}=0$}).
This leads to overlapping ranges, and some cases of the local velocity
u correspond to two or more reference velocities. In these cases, the
choice of \rcol{velocity shift $U_{\alpha}$} is dictated either by its history (to
avoid changes of reference) or by the cell's neighborhood (to reduce
the number of interface cells). Such enlarged ranges can only be
achieved with LBM models that are stable over a sufficiently large
velocity range.

%\rcol{As a side note, it can be useful to slightly shift the interval~(\ref{eq:criterion}) when the mean flow velocity falls close to a half-integer value $n \pm 0.5$ (as it is the case for several benchmark tests in Section~\ref{subsec:preliminary_investigations}).

\subsection{\label{subsec:shifting}Population reconstruction strategies at shift interfaces}

Now that criteria have been defined to allocate different velocity stencils to each part of the simulation domain, %the transfer of information at the transition interface must be specified. 
a proper algorithm for the transfer of information across the interface, between two areas with different stencils, must be defined.
Hereafter, all reconstruction strategies take place during the streaming step, which is considered hereafter in a `pull' manner (i.e., from the perspective of the cell receiving the streamed population) to ease our discussion without any loss of generality. In the %present context
conventional LB algorithm, a population can only be pulled from a location $\bm{x} + \bm{c}_{i\alpha}$ to the current node $\bm{x}$ if and only if it belongs to the same velocity discretization. Otherwize, the population is considered to be missing and a reconstruction step is then required, as already done, for example, in the context of boundary conditions. 

\subsubsection{\label{subsubsec:gauge_moments}Existing strategies}

%Looking at 
In first works on adaptive LBMs by Sun et al., missing populations are approximated by their equilibrium value~\cite{SUN_PRE_58_1998,SUN_PRE_61_2000,SUN_JCP_161_2000,SUN_PRE_68_2003,SUN_CF_33_2004}. 
%This has the advantage of being really cheap in terms of floating point operations, memory storage, and it can be applied to any kind of LBMs (quadrature-free or not) in a straightforward manner. Nonetheless, approximating missing populations by their equilibrium value must be balanced by
While it is computationally cheap and general, this method is limited by the fact that
a higher number of discrete velocities is required to recover the correct physics, which is in contradiction with our goal to propose efficient adaptive LBMs.

The above reconstruction strategy was recently improved by computing post-collision populations through their moment space~\cite{DORSCHNER_PRL_121_2018,ZAKIROV_KIAM_Preprint_Eng_2019}:
\begin{equation}\label{eq:reconstruction_moment}
   \mathcal{M}'\bm{h}^{*,'} = \mathcal{M}\bm{h}^{*}. 
\end{equation}
$\bm{h}^{*,'}$ is the vector of (missing) post-collision populations in the post-streaming velocity discretization, and 
\begin{equation}
\bm{h}^{*}=\bm{h} - \bm{\mathcal{M}^{-1}S\mathcal{M}} (\bm{h} - \bm{h}^{eq}) 
\end{equation}
is the vector of (known) post-collision populations in the pre-streaming velocity discretization. $\mathcal{M}$ and $\bm{S}$ are the moment and relaxation matrices in the pre-streaming velocity discretization, whereas $\mathcal{M}'$ is the moment matrix in the post-streaming one. This approach is the most expensive strategy to reconstruct missing populations since it requires the computation of \textit{all} moments of $h_i=f_i$ or $g_i$. Even if the size of the moment space remains small for the most basic velocity discretizations, such as D2Q9, the number of floating point operations required by this reconstruction grows as $V^2$, where $V$ is the size of the lattice. In addition, the full set of moments is only defined in an unique manner for tensor-product-based lattices (D2Q9, D2Q25, D2Q49, D2Q81, etc), whereas additional steps are required to derive full moment sets for more compact lattices (D2Q13, D2Q17, D2Q21, etc). 

One possible way to improve the efficiency would be to compute missing populations via their Hermite polynomial expansion, as already done, e.g., for hybrid solvers (DVM-LBM) in the context of rarefied gas flow simulations~\cite{ARISTOV_EUCASS_2017,ARISTOV_AIPCP_2132_2019,ARISTOV_JCS_40_2020}. This was very recently proposed through the adoption of a standard (non-recursive) regularized approach~\cite{LATT_MCS_72_2006a,CHEN_PA_362_2006} in the place of the reconstruction in the moment space~\cite{DORSCHNER_PRL_121_2018}. This allowed Zipunova et al. to reduce the computational overhead by one order of magnitude~\cite{ZIPUNOVA_DSFD_2020}. This latter result is explained by the fact that standard regularization steps discard high-order moments to reconstruct populations, hence reducing both the number of floating point operations and memory consumption at the same time -- as originally highlighted by Ladd and Verberg for standard LBMs~\cite{LADD_JSP_104_2001}. Even if the latter approach can be applied to more compact lattices, it is based on formulas derived in the context of quadrature-based LBMs, and consequently, it cannot be used, as it is, in the present quadrature-free formalism. 

\subsubsection{\label{subsubsec:gauge_CE}From Chapman-Enskog to Grad-type reconstructions}

In order not to restrict ourselves to tensor-product based lattices, %it seems appropriate to compute missing 
we compute post-collision populations through the decomposition 
\begin{equation}
    h_i^{*} = h_i^{eq} + (1-1/\tau_h) h_i^{neq},
\end{equation}
where the BGK operator is adopted for the sake of simplicity, and $\tau_h$ is the relaxation time corresponding to populations $h_i$. The interest reader may refer to Appendix~\ref{app:PrandtlExtension} for its extension to variable Prandtl numbers.

For (quadrature-free) LBMs based on numerical EDFs, while $h_i^{eq}$ can be computed through the root-finding algorithm with respect to the lattice of interest, $h_i^{neq}$ %is trickier to obtain. 
is not directly available.
One possible way to compute it is to rely on the CE expansion truncated at the first-order in Knudsen number~\cite{CHAPMAN_Book_3rd_1970}:
\begin{equation}\label{eq:reconstruction_CE}
    h_i^{neq} \approx h_i^{(1),\mathrm{CE}} = -\tau_h [\partial_t h_i^{eq} + \xi_{i\alpha} \partial_{\alpha} h_i^{eq}],
\end{equation}
 which is the most general form of regularization steps~\cite{LATT_MCS_72_2006a,CHEN_PA_362_2006,MALASPINAS_ARXIV_2015,COREIXAS_PRE_96_2017,MATTILA_PF_29_2017,WISSOCQ_ARXIV_2020_07353,ZIPUNOVA_DSFD_2020}, in the sense that it does not rely on any assumption regarding high-order moments or their relaxation frequency. Instead, it requires the computation of the space and time evolution of EDFs.
This is achieved by evaluating space and time derivatives through standard approaches such as finite-difference (FD) approximations. This may strongly impact the accuracy of the transition strategy because %small error evaluations 
errors at the level of populations might induce drastic discrepancies at the macroscopic level. Regarding the efficiency of this reconstruction, the computation of the time derivative requires the storage of equilibrium populations at previous time steps. Even in the most optimistic scenario (first-order Euler FD approximation), this strategy roughly doubles the memory consumption of the resulting LBM, leading to a decrease of parallel efficiency, especially in the context of GPU acceleration. 

%\subsubsection{\label{subsubsec:gauge_grad}Grad-type reconstruction}

The last strategy considered in this work originates from the kinetic theory of gases, and %more precisely, it 
is based on Grad's description of populations
\begin{equation}\label{eq:GradGeneral}
    h_i \approx h_i^{eq}+ h_i^{(1),\mathrm{Grad}} =h_i^{eq}(1+\phi_h)
\end{equation}
where $\phi_h$ is a deviation that accounts for (small) non-equilibrium contributions of populations $h_i$. These deviations can either be computed in terms of Hermite coefficients~\cite{GRADb_CPAM_2_1949,GRAD_CPAM_5_1952} or through the maximum entropy principle~\cite{KOGAN_JAMM_29_1965}. In both cases, the same form is obtained for monatomic gases:
\begin{equation}\label{eq:GradGeneral_f}
    \phi_f = \dfrac{ \sigma_{\alpha\beta}\overline{c}_{i\alpha}\overline{c}_{i\beta}}{2\rho T^2} + \dfrac{q_{\alpha}\overline{c}_{i\alpha}}{\rho C_p T^2}\left(\dfrac{\overline{c}_{i\chi}^2}{2T}-C_p\right), 
\end{equation}
where the traceless viscous stress tensor reads as
\begin{equation}
    -\sigma_{\alpha\beta}= \mu [\partial_{\alpha}u_\beta + \partial_{\beta}u_\alpha - (2/D) \partial_{\chi}u_{\chi}\delta_{\alpha\beta}]
\end{equation}
and Fourier's heat flux is
\begin{equation}
    q_{\alpha}= - \kappa \partial_{\alpha}T,
\end{equation}
with $\overline{c}_{i\alpha}=c_{i\alpha}-u_{\alpha}$ shifted peculiar discrete velocities.
$\mu = \rho \nu$, $\mu_b=(2/D-1/C_v)\mu$, $\nu$ and $\kappa$ are the dynamic viscosity, bulk viscosity, kinematic viscosity and thermal conductivity coefficients. Regarding the polyatomic non-equilibrium correction imposed through population $g_i$, one possible formulation is
\begin{equation}\label{eq:GradGeneral_g}
    \phi_g = 2 \dfrac{q_{\alpha}\overline{c}_{i\alpha}}{\rho C_p}. 
\end{equation}
The latter only impacts the definition of the heat flux% (see Appendix~\ref{app:transitionCE})
, and it naturally results from Rykov's model for polyatomic gases when non-elastic contributions are neglected~\cite{RYKOV_FD_10_1975}.

Similarly to the reconstruction based on the CE expansion~(\ref{eq:reconstruction_CE}), this methodology is lattice-independent, and consequently, more flexible than the reconstruction through the moment space~(\ref{eq:reconstruction_moment}). 
In addition, while the (numerical) equilibrium part $h_i^{eq}$ is computed via macroscopic quantities expressed in the lattice of interest, gradients are evaluated through standard finite-difference (FD) approximations. 
The reconstruction of missing information can then be done in an efficient and rather local manner, that is compliant with HPC architecture.

All of this makes Grad's approximation~(\ref{eq:GradGeneral}) a very appealing reconstruction technique from the theoretical and practical viewpoints. %, as opposed to its moment counterpart. (talk about the fact it filters some modes and is more stable than the first method? is it true?)} 
This strategy is therefore applied in the numerical section of the article.

\section{\label{sec:numerical_applications}Numerical applications and validation}

\subsection{\label{subsec:implementation}Implementation details}

\subsubsection{\label{subsubsec:underlying_LB_scheme}Underlying LB scheme}

In the context of LBMs, the DVBE~(\ref{eq:DVBE}) is numerically discretized using the very efficient collide-and-stream algorithm~(\ref{eq:LBE-BGK}), which can be divided into two consecutive steps:
\begin{subequations}
\begin{itemize}
    \item[-] Collision (here with the BGK approximation):
    \begin{equation}\label{eq:coll}
        h_{i}^*\left(\boldsymbol{x}, t\right) = h_{i}\left(\boldsymbol{x}, t\right) - \dfrac{1}{\tau_h} \left(h_{i}-h_{i}^{eq}\right)\left(\boldsymbol{x}, t\right),
    \end{equation}
    \item[-] Streaming (pull):
    \begin{equation}\label{eq:stream}
        h_{i}\left(\boldsymbol{x}, t + 1\right) = h_{i}^*\left(\boldsymbol{x}-\boldsymbol c_{i}, t\right),
    \end{equation}
\end{itemize}
\end{subequations}
where the latter step %is usually considered in its ``pull'' version for adaptive LBMs
has been formulated, as usual for adaptive LBMS, from a "pull" perspective.~\cite{SUN_PRE_58_1998,SUN_JCP_161_2000,SUN_PRE_61_2000,SUN_PRE_68_2003,SUN_CF_33_2004,DORSCHNER_PRL_121_2018,ZAKIROV_KIAM_Preprint_Eng_2019,ZIPUNOVA_DSFD_2020}.
Here, a double distribution function (DDF) formulation of LBMs is used ($h_i=f_i$ or $g_i$) to account for internal degrees of freedom of molecules via $g_i$, hence, leading to a flexible specific heat ratio $\gamma_r$~\cite{RYKOV_FD_10_1975,DUBROCA_ESAIM_10_2001,TITAREV_ECCOMAS_2006,NIE_PRE_77_2008,FRAPOLLI_PhD_2017,LATT_RSTA_378_2020}. 

%For the transition 
To allow transitions between shifted versions of the D2Q21 lattice, %numerical EDFs~(\ref{eq:exponential}) computed using eight constraints 
8-constraint numerical EDFs~(\ref{eq:feqNum_8Mom}) are considered in the rest of the paper (see Appendix~\ref{app:numericalEq} for more details). 
Interestingly, the equilibrium of the second set of populations ($g_i^{eq}$) is not computed via the root-finding solver, but instead, it is obtained from its monatomic counterpart ($f_i^{eq}$) through
\begin{equation}\label{eq:gEq}
    g_i^{eq} =(2C_v-D)T f_i^{eq},
\end{equation}
with the heat capacity at constant volume being $C_v=1/(\gamma_r - 1)$. Consequently, the polyatomic behavior is obtained at a low cost in terms of floating point operations, but %it doubles the memory storage requirements.
at the cost of doubled memory needs.

Due to the BGK approximation of the collision model, the above DDF-LBM is restricted to the simulation of flows with a unity Prandtl number, i.e., $\mathrm{Pr}=\rho C_p\nu/\kappa=1$. This can be corrected adopting a more advanced collision model, such as Shakhov's~\cite{SHAKHOV_FD_3_1968} or Ryjkov's~\cite{RYKOV_FD_10_1975}. Both can be deduced from Grad's formulation~(\ref{eq:GradGeneral}) by discarding the term related to stresses in the definition of $\phi_f$~(\ref{eq:GradGeneral_f}). More details about this extension can be found in Appendix~\ref{app:PrandtlExtension}. 
%For the present DDF-LBM, one ends up with
%\begin{subequations}\label{eq:Shakhov}
%\begin{widetext}
%\begin{align}
%    f_{i}(\boldsymbol{x}+\boldsymbol\xi_{i}, t + 1) &= \bigg[f_{i} \,-\, \dfrac{1}{\tau_f} (f_{i}-f_{i}^{eq}) + \bigg(\dfrac{1}{\tau_f}-\dfrac{1}{\tau_{\mathrm{Pr}}}\bigg) \dfrac{q_{\alpha}\overline{\xi}_{i\alpha}}{\rho C_p T^2}\bigg(\dfrac{\overline{\xi}_i^2}{2}-C_p\bigg)f_{i}^{eq}\bigg](\boldsymbol{x}, t),\\
%    g_{i}\left(\boldsymbol{x}+\boldsymbol\xi_{i}, t + 1\right) &= \bigg[g_{i} \,-\, \dfrac{1}{\tau_g} (g_{i}-g_{i}^{eq}) + \bigg(\dfrac{1}{\tau_g}-\dfrac{1}{\tau_{\mathrm{Pr}}}\bigg) \dfrac{q_{\alpha}\overline{\xi}_{i\alpha}}{\rho C_p}g_{i}^{eq}\bigg](\boldsymbol{x}, t),\label{eq:prandtlPoly}
%\end{align}
%\end{widetext}
%\end{subequations}
%with $\tau_f=\tau_g=0.5 + \nu/T$, and $\tau_{\mathrm{Pr}}=0.5 + (\nu/\mathrm{Pr})/T$.} %It is worth noting that since $g_i$ is related to internal degrees of freedom of molecules (trace of second-order central moment of $f_i$), one can simplify Eq.~(\ref{eq:prandtlPoly}) in
%%\begin{equation}
%%    g_{i}\left(\boldsymbol{x}+\boldsymbol\xi_{i}, t + 1\right) = \bigg[g_{i} \,-\, \dfrac{1}{\tau_{\mathrm{Pr}}} (g_{i}-g_{i}^{eq})\bigg](\boldsymbol{x}, t),\label{eq:prandtlPolySimp}
%%\end{equation}

Eventually, Grad's reconstruction technique~(\ref{eq:GradGeneral}) will also be used as an extension to the compressible case of (regularized) initial and boundary conditions~\cite{SKORDOS_PRE_49_1993,LATT_PRE_77_2008,DORSCHNER_JCP_295_2015}, %where the stress tensor $\sigma_{\alpha\beta}$ and the heat flux $q_{\alpha}$ will be computed using second-order FD approximations. 
using second-order FD approximations to approximate the viscous stress tensor $\sigma_{\alpha\beta}$ and the heat flux $q_{\alpha}$. For boundary conditions, it is worth noting that one could also rely on the bounce-back or the extrapolation of non-equilibrium populations for the computation of $\sigma_{\alpha\beta}$ and $q_{\alpha}$. Nevertheless, an extensive comparison of these approaches is out of the scope of this work.%, and consequently, it will be presented elsewhere. 

%\rcol{add a word about regularized collision model based on Grad's reconstruction?} 

\subsubsection{\label{subsubsec:stabilization}Non-linear stabilization technique}

By relying on numerical EDFs, the above collision models %are more stable than their polynomial counterpart. 
have been shown to be more stable than their polynomial counterparts for compressible flows~\cite{LATT_RSTA_378_2020}.
Yet, this is not sufficient to obtain stable simulations (1) in  the low-viscosity regime, (2) for severely under-resolved conditions, and (3) when strong compressibility effects (shock waves) are encountered. 
%\rcol{Following the best practices in the CFD community, one should rely on stabilization techniques that are able to (i) accurately capture sharp gradients and discontinuities without suffering from Gibbs oscillations% even in the vanishing viscosity limit
%, and (ii) take into account small-scale diffusive processes %(energy transfer from large scales to small ones through dissipation) 
%that occur at length scales that cannot possibly be resolved by the grid cell size~\cite{HIRSCH_Book_2007,GARNIER_Book_2009,PIROZZOLI_ARFM_43_2011}.
%To that end, the kinetic stabilization methodology proposed in our previous work is adopted, in order to tackle both points in a fairly good (but not perfect) manner~\cite{LATT_RSTA_378_2020}.} 
Following best practices in the CFD community, one should rely on stabilization techniques that are able to accurately capture sharp gradients and discontinuities without suffering from Gibbs oscillations while keeping smooth regions of the flow intact~\cite{HIRSCH_Book_2007,GARNIER_Book_2009,PIROZZOLI_ARFM_43_2011}.
To that end, the kinetic stabilization methodology proposed in our previous work is adopted, in order to tackle both points in a fairly good (but not perfect) manner~\cite{LATT_RSTA_378_2020}.
The latter consists in locally evaluating the departure from equilibrium through an approximation of the Knudsen number
\begin{equation}
    \epsilon_{\mathrm{Kn}} = \dfrac{1}{V}\sum_{i=0}^{V-1} \dfrac{\vert f_i-f_i^{eq}\vert}{f_i^{eq}},
\label{eq:DeltaAvg}
\end{equation}
which allows the distinction of all the features encountered during simulations. When $\epsilon_{\mathrm{Kn}}>0.01$, the recovery of the proper macroscopic behavior is at risk, and dissipation must be locally added to damp phenomena related to departures from equilibrium (e.g., shockwaves) or Gibb's oscillations induced by under-resolved conditions.
This simple yet powerful kinetic sensor is coupled with the common BGK collision operator through the dynamic relaxation time $\tau_f(\epsilon_{\mathrm{Kn}})= \tau_f \alpha(\epsilon_{\mathrm{Kn}})$. % following the same methodology as in our previous work~\cite{LATT_RSTA_378_2020}. 
\rcol{Interestingly, this stabilization mechanism can also be seen as a limiter for changes induced by the collision term. The interested reader may refer to the work by Gorban and Packwood (and therein references) for other types of non-equilibrium limiters~\cite{GORBAN_PA_414_2014}.}

In addition to its compliance with parallelism paradigms (local and fast evaluation), \rcol{this stabilization technique} was shown to lead to stable and accurate simulations of the inviscid Sod shock tube, and viscous flows past a 2D airfoil and 3D sphere in the supersonic regime~\cite{LATT_RSTA_378_2020}. 
Eventually, this stabilization technique barely depends on the chosen phase discretization, and is independent of the considered way of computing the equilibrium (polynomial or numerical). 
%It then seems to be 
It is therefore a good candidate to further increase the stability of the proposed adaptive LBMs when needed be. 
%\bcol{This will be the case for the simulation of non-linear phenomena that include discontinuities, such as the 1D and 2D Riemann problems (Section~\ref{subsec:further_application}).}

\subsubsection{\label{subsubsec:summary}Algorithm overview}

In summary, a time iteration of the proposed algorithm present itself as follows:
\begin{itemize}
    \item[-] Domain decomposition for the application of a local velocity space discretization (\ref{eq:criterion})
    \item[-] Computation of monatomic numerical EDFs $f_i^{eq}$~(\ref{eq:feqNum_8Mom}) based on the constraints~(\ref{eq:8moments})
    \item[-] Inclusion of polyatomic contributions to numerical EDFs through $g_i^{eq}$~(\ref{eq:gEq})
    \item[-] If needed, computation of Knudsen-dependent relaxation times based on the sensor~(\ref{eq:DeltaAvg})
    \item[-] Computation of post-collision populations~(\ref{eq:coll}) %based on the DDF formulation of Shakov's operator~(\ref{eq:Shakhov})
    \item[-] If the same lattice is used at $\bm{x}$ and \rcol{$\bm{x}-\bm{c}_i$} then the normal streaming step is applied~(\ref{eq:stream}). Otherwise, missing post-collision populations $h_i^*$ at $\bm{x}$ are reconstructed using Eq.~(\ref{eq:GradGeneral}), where macroscopic quantities and their gradients are evaluated at \rcol{$\bm{x}-\bm{c}_i$}.
\end{itemize}

\noindent It is interesting to point out that, the present adaptive approach barely modifies the standard collide-and-stream algorithm and, most importantly, does not require any space interpolation. %, hence, preserving the parallel efficiency of the numerical scheme. 
Extra steps consist in (1) the evaluation of the domain decomposition for the velocity discretization, and (2) reconstruction of missing post-collision populations when lattices at $\bm{x}$ and \rcol{$\bm{x}-\bm{c}_i$} differ.

The remainder of the paper is dedicated to the validation of the present approach, in terms of accuracy and robustness, for the simulation of linear and non-linear phenomena in the low-viscosity regime (Sections~\ref{subsec:preliminary_investigations} and~\ref{subsec:further_application} respectively). %\rcol{An investigation of (shear and thermal) diffusivity properties is also provided in Appendix~\ref{app:decay} for the sake of completeness.} 

\subsection{\label{subsec:preliminary_investigations}Preliminary investigations: Shifting impact on linear properties}

%Since the macroscopic behavior of LBMs based on numerical EDFs cannot be \textit{exactly} evaluated through standard asymptotic expansions, it is proposed to assess the operability range of these LBMs through the evaluation of transport coefficients (kinematic viscosity and thermal diffusivity) as well as the speed of sound. This is done by quantifying the impact of numerous parameters (number of constraints, Mach number, reference temperature and velocity shift) on the decay of both shear and thermal waves, as well as, the propagation of an acoustic perturbation.   

It is first proposed to investigate the linear behavior of the present approach for a wide range of flow velocities. %how changing the reference frame (in either a static or dynamic manner) does impact the physical properties of the corresponding LBMs. 
Usually, this is done via linear stability analyses~\cite{HIRSCH_Book_2007,HOSSEINI_PRE_100_2019},% but in the present case, this is not possible 
which can however not been carried out presently due to the numerical nature of the EDF. Alternatively, one can directly simulate the space-time evolution  
of small perturbations (shear, thermal and acoustic waves) that are superimposed to a uniform flow. 

Hereafter, our adaptive formulation will be validated using the D2Q21 lattice based on the numerical EDF computed through $M=8$ constraints (\ref{eq:G0_8mom})-(\ref{eq:G3yaa_8mom}). For all considered tests, the heat capacity ratio, the reference temperature and density will be fixed to $\gamma_r=1.4$, $T_0=0.7$ and $\rho_0=1$. All of this is done in the inviscid context ($\nu=0$) to further highlight the accuracy and stability of the present approach. If not otherwise stated, reference data are obtained using the D2Q49 lattice based on the numerical EDF computed through $M=13$ constraints (\ref{eq:G0})-(\ref{eq:G3yyy}), which exactly recovers the NSF equations, and with the same parameters $(\gamma_r,T_0)=(1.4,0.7)$. 

%\rcol{Add time evolution of the mass to confirm the conservativity of the present approach? Add something about variable Prandtl?}

%\rcol{If not otherwise stated, the kinetic stabilization technique will be used for all numerical simulations.}

Due to the linear nature of the following flow configurations, the lattice self adjusts at the beginning of the simulation and does not change %after that. 
afterwards. It then restricts the reconstruction~(\ref{eq:GradGeneral})-(\ref{eq:GradGeneral_g}) to the initialization step, %hence, highlighting 
which highlights its interesting properties as an extended initial condition. All boundary conditions are periodic.

\subsubsection{\label{subsubsec:shear_wave}Shear wave}

\begin{figure}[htbp!]
	\centering
\includegraphics[width=.45\textwidth]{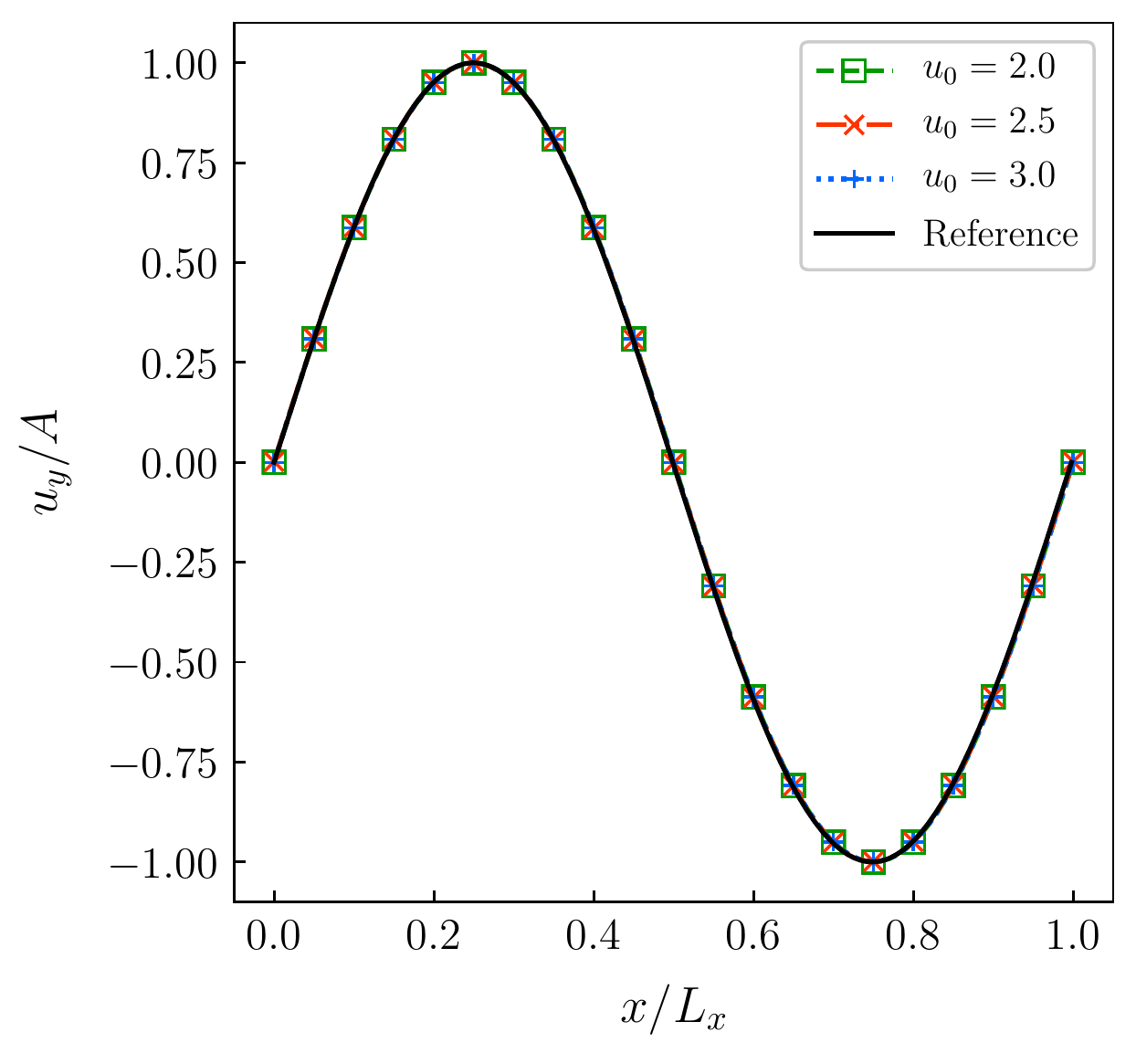}
\caption{Convection of a shear wave in an inviscid flow with a grid mesh composed of $L_x=100$ points. Normalized transverse velocity fluctuations $u_y/A$ are plotted after $t=15 t_c$ for three different flow velocities $u_0=2$, $2.5$ and $3$. Reference results correspond to the initial state for $u_0=0$ and $L_x=500$ points.}
\label{Fig:shearWave}
\end{figure}

\begin{figure*}[htbp!]
	\centering
\includegraphics[width=.99\textwidth]{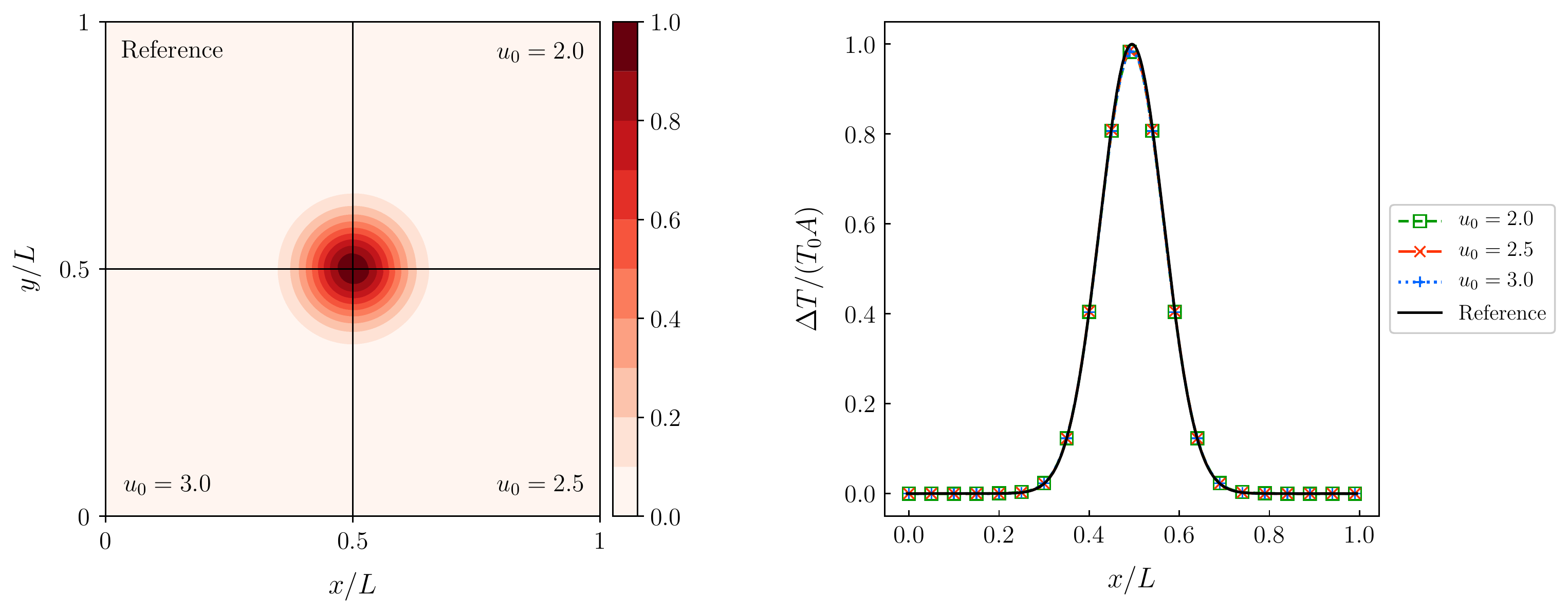}
\caption{Convection of an entropy spot in an inviscid flow with a grid mesh composed of $L=100$ points in each direction: global view and slice along the $x$-axis (from left to right). Normalized temperature fluctuations $\Delta T/(AT_0)$ are plotted at $t=15 t_c$ for three different flow velocities $u_0=2$, $2.5$ and $3$. Reference results correspond to the initial state based on a mesh composed of $L=100$ and $500$ points for the global view and the slice respectively. The numerical dissipation with respect to the theoretical temperature peak is $1.2\%$, $0.9\%$ and $1.2\%$ for $u_0=2$, $2.5$ and $3$ respectively.}
\label{Fig:entropySpot}
\end{figure*}

%The linear behavior of the present approach is investigated by superimposing small perturbations %$(\rho',\bm{u}',T')$ 
%to an uniform mean flow. %$(\rho_0,\bm{u}_0,T_0)$. 
%Here, we focus the study on the time evolution of (1) a shear and (2) a thermal wave. Corresponding initial states were proposed, e.g., by Shan and Chen for the derivation of a multi-relaxation-time collision model~\cite{SHAN_IJMPC_18_2007}. The shear wave is defined by
%\begin{equation}
   % \rho = \rho_0,\: u_x = u_0, \: u_y = A \sin(2\pi x / L_x), \: T=T_0, 
%\end{equation}
%whereas the thermal wave is initialized by imposing
%\begin{equation}
%   \rho = \rho_0+A\sin(2\pi x / L_x),\: u_x = u_0, \: u_y = 0, \: T=\rho_0 T_0/\rho. 
%\end{equation}
%For both configurations, the perturbations' amplitude are $A=0.001$, and the mean quantities $(\rho_0,T_0) = (1,0.6)$. 
%The mean velocity $u_0$ will span a large range of values by imposing $u_0=U+\Delta u$ with $U$ the velocity shift and $\Delta u\in\{0, 0.2, 0.4, 0.6, 0.8, 1\}$. 

We first investigate the transport of a shear wave by an inviscid ($\nu=0$) uniform mean flow. This wave corresponds to a small sinusoidal perturbation (in the transverse velocity field) that is superimposed to a uniform mean flow~\cite{SHAN_PRE_100_2019,RENARD_ARXIV_2020_03644}:
\begin{equation}
    \rho = \rho_0,\: u_x = u_0, \: u_y = A \sin(2\pi x / L_x), \: T=T_0, 
\end{equation}
where the perturbations' amplitude is $A=0.001$, the mean quantities being $(\rho_0,T_0) = (1,0.7)$, and the flow velocity $u_0$ is a parameter that is varied to evaluate the accuracy and robustness of the present approach in high-speed flow conditions. The simulation domain is quasi-one-dimensional, i.e. $[L_x\times L_y] = [100\times 1]$, where $L_{\alpha}$ is the number of points in direction $\alpha=x$ or $y$. 

In order to quantify the numerical dissipation and dispersion of the proposed approach, it is mandatory to \textit{exactly} propagate the shear wave over the same distance %whatever the
for any value of the mean flow velocity $u_0$. This %can be
is done by carefully choosing $u_0$ and the time ($t=N t_c$) at which results are compared, where the characteristic time is defined as $t_c=L_x/u_0$, and $N\in \mathbb{N}^*$. By taking $u_0\in \{2.0,2.5,3.0\}$, it is possible to compare all results at $t=15 t_c$, which corresponds to an integer number of iterations for all values of $u_0$ (i.e., $t = 750$, $600$ and $500$ time iterations respectively).

Results are compared with a reference solution in Fig.~\ref{Fig:shearWave}. The overlap between simulation and reference data prove the excellent spectral properties (dispersion and dissipation of shear waves) of the present approach for the considered grid mesh $L_x=100$. Quantitatively speaking, the numerical dissipation with respect to the reference velocity peak 
\begin{equation}
\nu_{\mathrm{num}}=\dfrac{\vert \max(u_{y}^{\mathrm{ref}})-\max(u_y) \vert}{\max(u_{y}^{\mathrm{ref}})},
\end{equation}
is $0.3\%$, $0.9\%$ and $0.7\%$ for $u_0=2$, $2.5$ and $3$ respectively. %This confirms the very good numerical properties of our approach for the propagation of shear waves.

%This inviscid study is supplemented with kinematic viscosity measurements for various values of $\nu=10^{-a}$. 

\subsubsection{\label{subsubsec:entropy_spot}Entropy spot}

The linear properties of our approach are further investigated through the transport of an (inviscid) entropy spot. The latter is initialized as a (Gaussian) hot spot which is superimposed to a uniform velocity mean flow at constant pressure. Such an initial state was proposed by Fabre et al.~\cite{FABRE_PoF_13_2001}, and recently investigated in the context of hybrid LBMs~\cite{FARAG_PoF_32_2020}:
\begin{equation}
    \rho = \rho_0[1 - A\exp(-r^2)], \: T=T_0[1 + A\exp(-r^2)], 
\end{equation}
with $u_x = u_0$, $u_y=0$, $\rho_0=1$, $T_0=0.7$, $A=0.001$, and $r^2=[(x-x_c)^2+(y-y_c)^2]/R^2$ where $(x_c,y_c)$ are the coordinates of the hot spot center, $R=L/10$ is related to the spot width, and $L$ is the characteristic length of the simulation domain. In addition to the evaluation of the spectral properties (dispersion and dissipation) related to thermal waves, this benchmark test allows us to further quantify possible isotropy issues induced by our approach.

Here, the entropy spot is convected using $u_0=2$, $2.5$ and $3$, the latter allowing a proper comparison of the temperature fields at $t=15 t_c$, as for the previous testcase. Results are compiled in Fig.~\ref{Fig:entropySpot} for a relatively coarse mesh: $L_x=L_y=L=100$ points, leading to a full width at half height of $2R = 20$ points. All configurations prove that the present approach can transport small thermal fluctuations over long distances at the correct speed (negligible dispersion error), with only little loss of information (about $1\%$ of the peak amplitude for all configurations), 
%without introducing too much anisotropy through dispersion issues
while keeping the isotropic nature of the temperature field.

\subsubsection{\label{subsubsec:acoustic_pulse}Acoustic pulse}

\begin{figure*}[hbtp!]
	\centering
\includegraphics[width=.99\textwidth]{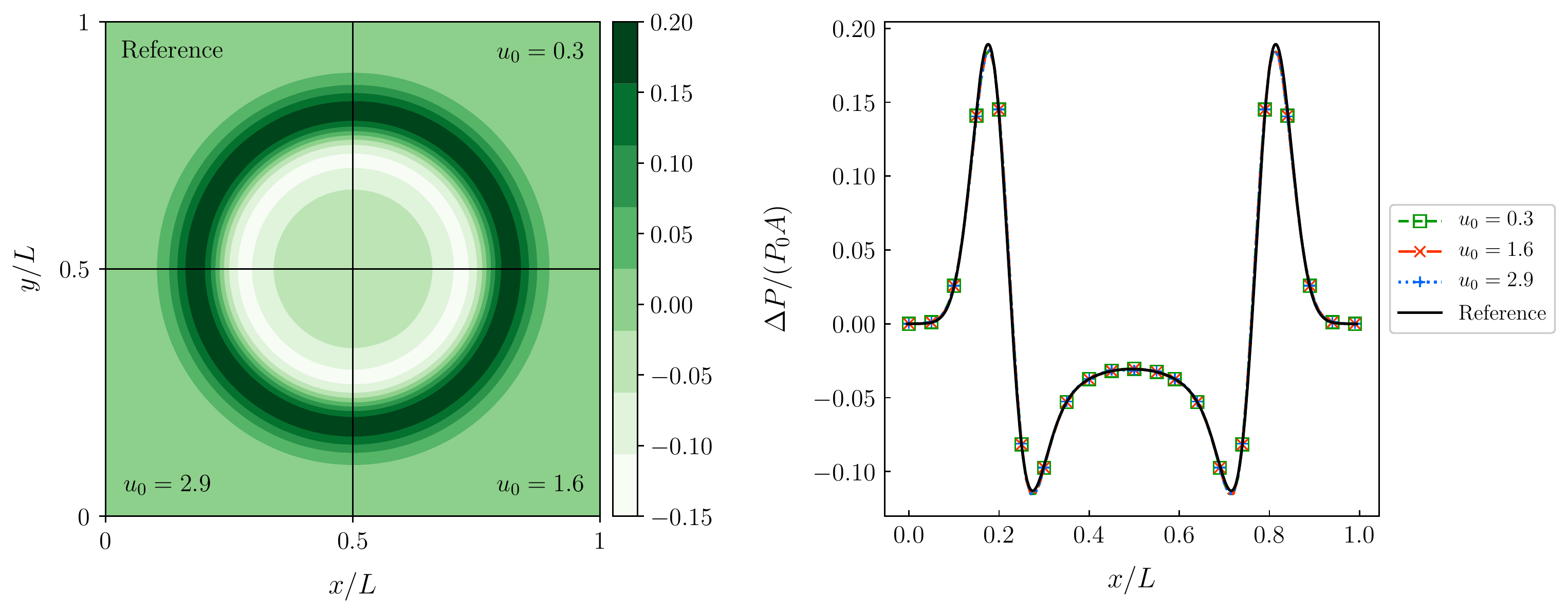}
\caption{Convection of an acoustic pulse in an inviscid flow with a grid mesh composed of $L=100$ points in each direction: global view (left) and slice along the $x$-axis (right). Normalized pressure fluctuations $\Delta P/ (A P_0)$ are plotted after $30$ time iterations for four different flow velocities $u_0=0$, $1.3$, $2.6$ and $3.9$. Data corresponding to the latter three simulation has been recentered to ease to comparison with the reference state. Reference results correspond to $u_0=0$, and are based on a mesh composed of $L=100$ and $500$ points for the global view and the slice respectively.}
\label{Fig:acoustic_pulse}
\end{figure*}

%\subsubsection{\label{sec:vortex_transport}Vortex transport by a uniform flow}
%This benchmark test is used to investigate the ability of the proposed approach to transport information over long distances without loss.
The last of these linear testcases deals with the generation and propagation of (isentropic) acoustic waves. The latter are induced by a (small) pressure disturbance that is superimposed to a mean velocity flow field:
\begin{subequations}
\begin{equation}
    P = P_0[1 + A\exp(-r^2)],\: u_x = u_0,\:  u_y=0 
\end{equation}
where the density and temperature flow fields are computed via Laplace's law, i.e.,
\begin{equation}
    \rho = \rho_0(P/P_0)^{1/\gamma_r} ,\: T = T_0(\rho/\rho_0)^{\gamma_r-1}, 
\end{equation}
with $P_0 = \rho_0 T_0$.
\end{subequations}
In addition to being convected by the mean flow velocity, this perturbation evolves over time and space. Hence, the comparison of the pressure field at the same physical time for various values of $u_0$ is not as straightforward as before. We then propose to recenter the pressure field around ($x_c$,$y_c$) after a fixed number of iterations, so that, we will still be able to quantify numerical dissipation and dispersion of the proposed approach. 

Corresponding results are plotted in Fig.~\ref{Fig:acoustic_pulse}, for three mean velocities ($u_0= 0.3$, $1.6$ and $2.9$). This is done after 30 iterations which corresponds to $t/t_c\approx 0.297$ where the acoustic characteristic time is defined as $t_c=L/\sqrt{\gamma_r T_0}$ and $L_x=L_y=L=100$ points. The different pressure flow fields confirm the very good isotropic dispersion properties of our approach, and the pressure profiles further highlight its low numerical dissipation property.

\subsection{\label{subsec:further_application}Further validation: Shifting impact on non-linear phenomena}

In the above section, it was shown that, in the linear regime, it is possible to extend the stability domain of our compressible DDF-LBM by adjusting the lattice to the local velocity field. To further validate the adaptive reconstruction of missing post-collision populations, we then propose to move towards more complex testcases, that include non-linearities, such as shock waves.

In the following, the kinetic sensor~(\ref{eq:DeltaAvg}) is used to reduce the generation of spurious Gibbs oscillations induced by discontinuities, and which are more prominent in under-resolved conditions and/or in the inviscid regime. All initial and boundary conditions are based on Grad's reconstruction~(\ref{eq:GradGeneral})-(\ref{eq:GradGeneral_g}), where second-order FD approximations are used for the computation of $\sigma_{\alpha\beta}$ and $q_{\alpha}$. 

\begin{figure*}[btp!]
	\centering
\includegraphics[width=.9\textwidth]{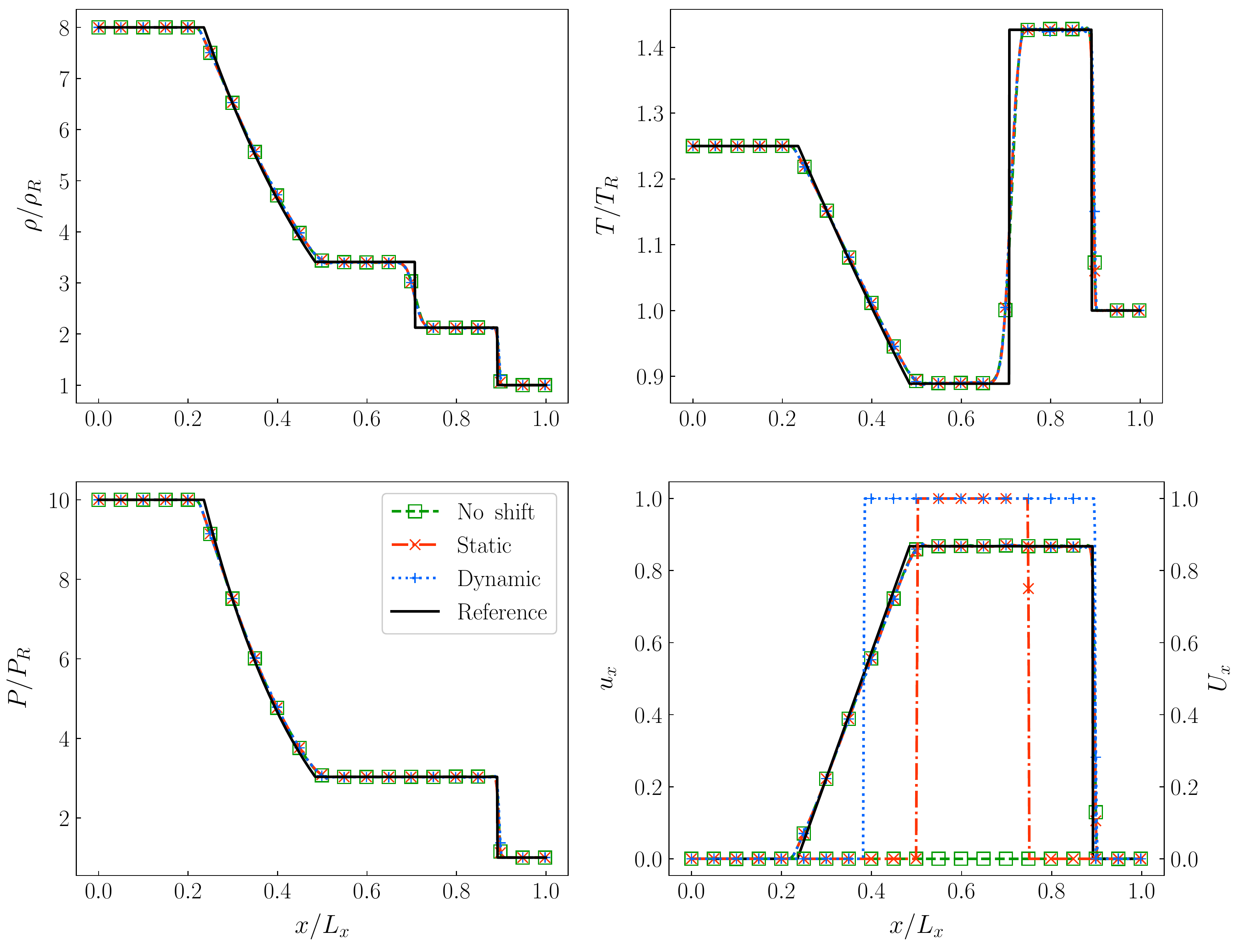}
\caption{Comparison of different fields (density $\rho$, temperature $T$, pressure $P$, velocity $u_x$) against exact solution of the Sod shock tube configuration. The spatial evolution of the shift ($U_x$) is superimposed to the velocity field $u_x$ The two configurations `No Shift' and `Static' corresponds to a fixed shift $U_x=0$ for $0\leq x\leq L$, and $U_x=1$ for $L/2 \leq x\leq 3L/4$ respectively. The `Dynamic' configuration corresponds to our adaptive approach based on the criterion~(\ref{eq:criterion}) with $n_0 = 0.51$. %: (black plain lines) exact solution, (green dahsed lines) LB simulation with no shift, (red dotted dashed lines) LB simulation with a static shift and (blue dotted lines) LB simulation with adaptive shifts. The shifts corresponding to each simulation are shown with dashed/dotted lines of the corresponding color.
}
\label{Fig:sod1_data}
\end{figure*}

\begin{figure*}[btp!]
	\centering
\includegraphics[width=.95\textwidth]{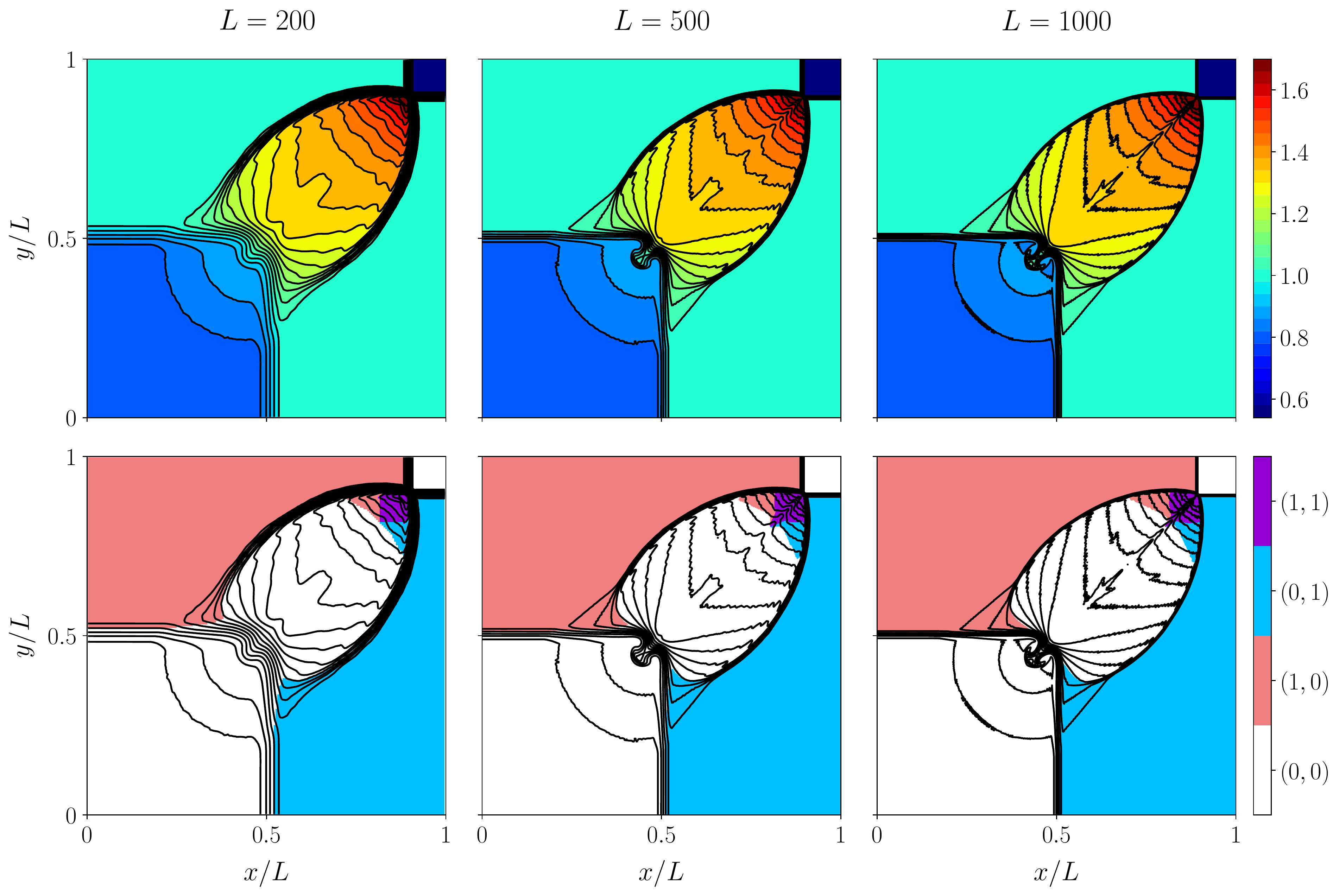}
\caption{2D Riemann configuration: density $\rho$ and velocity shift fields $U_x, U_y$ (from top to bottom). Black plain lines are density iso-contours: 30 equally-spaced levels in the density interval $\rho\in[0.54\text{ }1.7]$ following the instructions provided by Schulz-Rinne et al.~\cite{schulz1993numerical}. The most common features of the flow match results obtained by previous works~\cite{schulz1993numerical,LAX_SIAM_19_1998,KURGANOV_NMPDE_18_2002}. For the sake of clarity, only positive velocity shifts are plotted.}
\label{Fig:riemann2d}
\end{figure*}

\subsubsection{\label{subsec:sod_tube}Sod shock tube}
To assess the ability of the proposed solver to deal with highly compressible phenomena, we first consider a 1-D Riemann problem commonly referred to as Sod's shock tube~\cite{sod1978survey}. In its most popular form, it consists of a 1-D simulation domain initially divided into two sub-domains with different densities and temperatures, {i.e.},
\begin{equation}
(\rho_L/\rho_R,T_L/T_R)=(8,1.25)    
\end{equation}
with $(\rho_R,T_R)=(1,T_0)$ and $u_x=u_y=0$. 

For this particular configuration, the discontinuities in the macroscopic fields lead to the generation and propagation of three different waves, namely, the rarefaction wave, the contact discontinuity and the shock wave~\cite{sod1978survey}. 
The rarefaction wave propagates towards the high-density region of the simulation domain and induces smooth variations of density, temperature and velocity fields. 
On the contrary, both the contact discontinuity and the shock wave propagate towards the low-density region and lead to discontinuous macroscopic fields, with the exception of the (normal) velocity and pressure fields which remain constant at the contact discontinuity. 

Hereafter, we investigate the ability of our adaptive LBM to accurately reproduce the above features in the vanishing viscosity limit ($\nu=0$), using a quasi-1D simulation domain $[L_x \times L_y]=[500 \times 1]$. More precisely, we focus our attention on the impact of static and dynamic phase space transitions on discontinuities. Corresponding results are complied in Fig.~\ref{Fig:sod1_data} for three different configurations: (1) no shift $U_x=0$, static shift $U_x=1$ for $x\in [L/2,3L/4]$, and (3) dynamic shift based on our criterion~(\ref{eq:criterion}).% with $n_0=0.51$. 

Interestingly, all waves are properly generated and correctly evolve over time and space. 
The static configuration proves that discontinuities can travel through phase space transitions without generating high-amplitude spurious waves. 
This validates our reconstruction strategy~(\ref{eq:GradGeneral})-(\ref{eq:GradGeneral_g}). 
In addition, the dynamic configuration shows how the transition follows the velocity field, as well as, its (almost) negligible impact on the macroscopic flow fields. 
It is also worth noting that boundary conditions based on Grad's reconstruction~(\ref{eq:GradGeneral})-(\ref{eq:GradGeneral_g}) do not generate spurious oscillations and lead to accurate results. 
Eventually, only the contact discontinuity is overdissipated. 
This can be easily corrected by either increasing the mesh resolution, or by fine tuning our sensor --as usually done for LBMs based on shock-capturing techniques~\cite{COREIXAS_PhD_2018}.

\subsubsection{\label{subsec:2D_Riemann}Riemann 2D configuration}

\begin{table}[btp!]
\renewcommand{\arraystretch}{1.4}%
\begin{tabularx}{\columnwidth}{>{\centering\arraybackslash}X >{\centering\arraybackslash}X}
\toprule\toprule
   $\rho^{[2]} = 1$ & $\rho^{[1]} = 0.513$\\ 
   $u_x^{[2]} = 0.7276$ & $u_x^{[1]} = 0$\\ 
   $u_y^{[2]} = 0$ & $u_y^{[1]} = 0$\\
   $P^{[2]} = 1$ & $P^{[1]} = 0.4$\\ 
   \midrule
   $\rho^{[3]} = 0.8$ & $\rho^{[4]} = 1$\\ 
   $u_x^{[3]} = 0$ & $u_x^{[4]} = 0$\\ 
   $u_y^{[3]} = 0$ & $u_y^{[4]} = 0.7276$\\
   $P^{[3]} = 1$ & $P^{[4]} = 1$\\
  \bottomrule\bottomrule
\end{tabularx}
\caption{Initialisation of each quadrant $[q]$ of the simulation domain ($q\in \llbracket 1, 4 \rrbracket$). This setup was proposed in previous works, and it corresponds to configurations F~\cite{schulz1993numerical}, and 12~\cite{LAX_SIAM_19_1998,KURGANOV_NMPDE_18_2002} respectively.}
\label{Tab:Riemann2D}
\end{table}

While the ability of our adaptive approach to cope with non-linearities has been proven in the 1D case, hereafter, we further intend to validate it against a 2D Riemann problem. The latter corresponds to four 1D Riemann problems (or Sod shock tubes) that are initialized by dividing the simulation domain into four quadrants. Depending on the density, pressure and velocity conditions imposed for each quadrant, a large panel of compressible phenomena arise at quadrant interfaces, and especially in the vicinity of the simulation domain center. 

These phenomena were extensively studied in several seminal works (Schulz-Rinne et al.~\cite{schulz1993numerical}, Lax and Lui~\cite{LAX_SIAM_19_1998}, Kurganov and Tadmor~\cite{KURGANOV_NMPDE_18_2002}, etc). 
Based on the latter works, we focus our attention on the configuration depicted in Table.~\ref{Tab:Riemann2D}. This benchmark test is of particular interest to validate adaptive transitions of phase space due to the complex interplay occurring between all waves.

Generally speaking, several features/phenomena are expected when simulating this particular 2D Riemann problem. 
First, the initial state is symmetric with respect to the diagonal axis ($x=y$), hence, it can be supposed that macroscopic fields will keep this symmetry property. 
Second, quadrants [2] and [4] show velocities ($u_x$ and $u_y$ respectively) higher than $0.5$, whereas other quadrants ([1] and [3]) are initialized with a flow at rest. Consequently, it is guaranteed that at least three versions of the D2Q21 lattice will coexist in the following simulation: $(U_x,U_y)=(0,0)$, $(1,0)$, and $(0,1)$. 
Third, all quadrants but [1] share the same pressure field, and have identical perpendicular velocities at the quadrant interfaces, which implies that contact discontinuities are to be expected at these interfaces. 
Eventually, more complex calculations show that two shock waves form at interfaces with quadrant [1], and they propagate towards the upper right corner of the simulation domain. Doing so, they interact with each other, and generate a complex pattern in this part of the simulation domain. The latter pattern further forces contact discontinuities to roll up into a pair of vortices inside the third quadrant~\cite{schulz1993numerical,LAX_SIAM_19_1998,KURGANOV_NMPDE_18_2002}.  

Hereafter, the above Riemann 2D problem is simulated in a domain $[L_x \times L_y]$ using three grid mesh resolution: $L_x=L_y=L=200$, $500$, and $1000$. The coarsest and finest meshes are used to quantify the robustness of our approach in under- and well-resolved conditions, whereas the intermediate case is typical of mesh resolution encountered in the literature~\cite{WILDE_PRE_101_2020}. The post-processing proposed in the latter work is also adopted hereafter, i.e., only half of the simulation domain is plotted. Density $\rho$ and velocity shifts $(U_x,U_y)$ fields are compiled in Fig.~\ref{Fig:riemann2d} for the three grid mesh resolutions. 

Interestingly, the above-mentioned features are well recovered (symmetry, complex flow patterns), even if contact discontinuities and the related `mushroom'-shaped dipole are generally over-dissipated --as it was already the case for the Sod chock tube. 
Nevertheless, by increasing the mesh resolution, finer structures appear close to $(x,y)=(L/2,L/2)$ and inside the complex pattern induced by the interplay between the two shock waves. 
All of this favorably compares with previous studies~\cite{schulz1993numerical,LAX_SIAM_19_1998,KURGANOV_NMPDE_18_2002,WILDE_PRE_101_2020}, even if high-frequency oscillations are visible in the first quadrant. \rcol{Since the present approach shows  low numerical dissipation levels (as proved by results presented in Section~\ref{subsec:preliminary_investigations}), this suggests that our stabilization strategy is too naive. It would then benefit from fine tuning (i) to better damp these high-frequency waves, as well as, (ii) to decrease the unnecessary amount of artificial viscosity that impacts the growth of the `mushroom'-shaped dipole}. 

Regarding phase space transitions, it is interesting to note that they accurately follow the shock waves --and the complex pattern generated by the latter-- in their propagation towards the upper right side of the simulation domain. 
Velocity flow conditions also generate a new phase space, which is based on the shift $(1,1)$, upstream the intersection between the two shock waves. Even though they are not plotted for the sake of clarity, it is worth noting that three negative shifts also appear inside the dipole when the mesh resolution is increased: $(U_x,U_y)=(-1,0)$, $(0,-1)$, and $(-1,-1)$. %\bcol{Add a video as supplementary material to show them?}

These simulation results confirm the viability of our approach for interpolation-free adaptive LBMs in the context of high-speed compressible flow simulations.

\section{Conclusion\label{sec:conclusion}}

Since the introduction of adaptive stencils in the LB context, %more than twenty years ago, 
no interpolation-free formulation has been proposed in the literature. Knowing that space interpolations are computationally expensive, potentially prone to anisotropy issues (depending on the considered stencil for interpolation points), and deteriorate the parallel efficiency of the solver, \rcol{it seems appropriate to propose a solution to this issue. }%\hl{it is rather surprising that such an issue has never been addressed before.} 
In this context, the present work aims at proving the viability of interpolation-free adaptive LBMs for high-speed compressible flow simulations. 

The latter formulation leads to a series of new restrictions and challenges to realize the transition between two phase spaces in a stable and accurate manner. In the context of LBMs based on polynomial equilibrium distribution functions (EDFs), we investigated both points through linear stability analyses of tensor-product based lattices. 
As a striking result, the overlap of stability domains is only possible for DdQq$^d$ lattices with $q\geq 9$ (i.e., $\xi_{i\alpha} \in \{0,\pm1,\pm2,\pm3,\pm4\}$). Obviously, this type of adaptive LBMs has only little interest in practice. 
Alternatively, LBMs based on numerical EDFs are more flexible due to their quadrature-free nature, and they show interesting stability properties for more compact lattices, and notably, the D2Q21 lattice.

To couple the two shifted versions of the solver, one must answer the following questions: (1) when to switch between the two solvers, and (2) how to exchange the information at the interface between them. 
To tackle the former issue, we explored a number of solutions based on either already existing, or new switching criteria. 
In the end, it is proposed to locally adjust the lattice depending on velocity fluctuations, %and \rcol{temperature fluctuations}
in accordance with the original formulation of adaptive LBMs, and stability ranges of numerical EDFs. 
Regarding the transfer of information at phase space transitions, missing populations are recontructed through their equilibrium and non-equilibrium contributions. 
While the EDF can be easily computed for both polynomial and numerical approaches, the non-equilibrium part cannot be obtained in a straightforward manner for the latter case. 
Two (regularized) approaches are proposed to compute it, i.e., Chapman-Enskog's and Grad's formulations.
The former relies on the space-time evolution of EDFs, and requires their storage at previous time step(s), which is not suitable from the point of view of memory consumption. % \bcol{Add an appendix where we compare both approaches from the accuracy viewpoint? for the 2D Riemann problem?}. 
On the contrary, Grad's approach is a generalization of already existing regularized approaches which here assumes that non-equilibrium contributions are proportional to EDFs and diffusive fluxes (viscous stress tensor and heat flux). Ultimately, this provides an efficient, on-the-fly reconstruction strategy that can be applied to both polynomial and numerical EDFs. %In addition, we further rely on this extended strategy in order to account for non-equilibrium contributions for initial and boundary conditions.   

To account for internal degrees of freedom --mandatory for the simulation of polyatomic gas flows-- the above interpolation-free strategy is build on top of a double-distribution-function formulation (DDF-LBM).
Following the methodology provided in~\cite{LATT_RSTA_378_2020}, the D2Q21 lattice is coupled with a numerical EDF based on all convective constraints (8 moments in 2D) in order to obtain a purely LB solver that offers a good trade-off between accuracy, efficiency and stability. %\rcol{In order to make the Prandtl number easily adjustable, a DDF extension of Shakov's collision model is also proposed.} 
In case better stability is required, we further propose to locally adjust the kinematic viscosity depending on the departure of populations from their equilibrium state, which shares similarities with shock capturing techniques \rcol{and flux limiters}.
%\hl{Ali: and entropic fomulation?}
%In practice, it increases viscosity in under-resolved parts of the simulation domain, close to steep gradients such as discontinuities (slip lines, shock waves), and acts fairly well as subgrid scale models and/or shock capturing techniques. 

%The greatest restriction imposed by the phase space transition is to find the proper combination lattice-equilibrium to ensure .   

%All of this proves it is possible to design interpolation-free adaptive LBMs for the simulation of high-speed compressible flows. 

%In the end, the very good results obtained with our reconstruction strategy confirms that all moments of populations are not required to build accurate and stable phase space transitions even in the presence of discontinuities. Not only this drastically improves the efficiency of the reconstruction step, but it might even lead to more stable simulations. The latter point is under investigation and corresponding results will be presented in a future work.

The above DDF-LBM is validated in the low viscosity regime through several benchmark tests of increasing complexity. 
The first three tests aim at assessing the accuracy and stability of the present approach in the linear regime through the transport of shear, thermal and acoustic disturbances. 
They confirm the excellent numerical properties (low dispersion and dissipation) of our approach even for very high flow velocities ($\vert\vert \bm{u}\vert\vert\leq 3$). 
The last two are far more challenging as they include the generation and propagation of strong non-linear phenomena. 
The first one is based on Sod shock tube test, and it is used to check the ability of our adaptive strategy to handle discontinuities in both a static and dynamic manner. The last test consists in a 2D Riemann problem that involves a complex interplay of waves, and which is perfect to assess the dynamic behavior of our interpolation-free strategy. This test shows that our adaptive approach is also able to accurately handle non-linearities and patterns induced by their interplay.

%In summary, we proposed an interpolation-free formulation of adaptive LBMs that is compatible with any kind of lattice and equilibrium type. With only 21 discrete velocities, the latter approach can be used for the simulation of high-speed compressible and polyatomic gas flows.

Interestingly, we observed that our reconstruction strategy can also be used as extended initial and boundary conditions. As future work, in-depth comparisons should be conducted in order to quantify their benefits and limitations as compared to already existing methods. Eventually, the reconstruction of missing populations on interfaces can also be used to propose new LB collision models. In this case, pre-collision populations are replaced by their regularized counterpart on each cell and at every time step. This idea will be explored in future works.

\begin{acknowledgments}
The authors received no financial support for the research, authorship, and/or publication of this article. 
%S.A.H. would like to acknowledge the financial support of the Deutsche Forschungsgemeinschaft (DFG, German Research Foundation) in TRR 287 (Project-ID 422037413).
\end{acknowledgments}

\section*{Data availability}
The data that support the findings of this study are available
from the corresponding author upon reasonable request.

\appendix

%\onecolumngrid
%\begin{widetext}

{\color{black}{
\section{Approximation of the Maxwellian in a given reference frame\label{app:maxwell_hermite}}

\begin{figure*}[bt]
\centering
\includegraphics[width=0.99\textwidth]{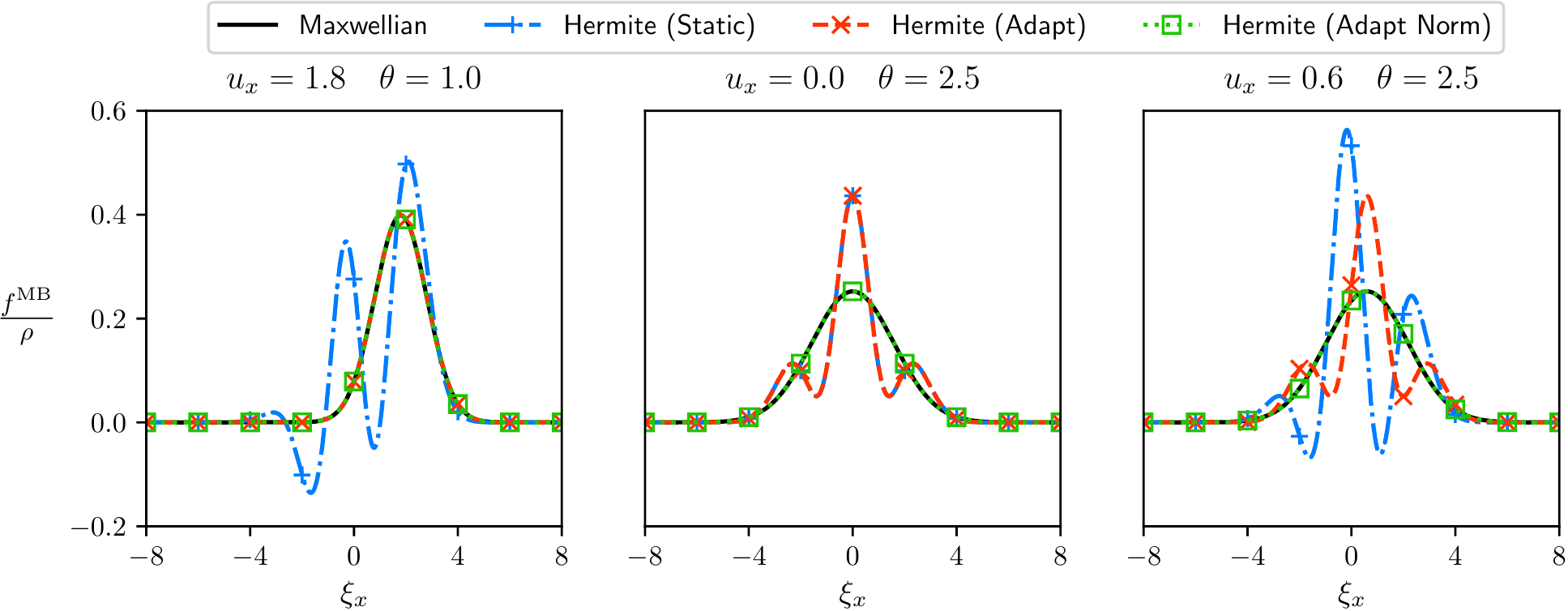}
\caption{Comparison of the Maxwellian with three Hermite polynomial expansion truncated at order $4$. The static, adaptive and normalized adaptive expansions are based on Hermite, central Hermite, and $\theta$-normalized central Hermite moments respectively. Only the latter set of moments lead to an error-free approximation of the Maxwellian, whereas the adaptive configuration shows $\theta$-dependent errors. The static configuration suffers from both $u$- and $\theta$-dependent errors.}
\label{fig:maxwell_vs_hermite}    
\end{figure*}

In the seminal work by Grad~\cite{GRADb_CPAM_2_1949}, velocity distribution functions are quantities that propagate at a temperature-normalized peculiar velocity
\begin{equation}\label{eq:cont_norm_peculiar}
    \xi_{\alpha} = (c_{\alpha} - u_{\alpha})/\sqrt{\theta} \quad\Longleftrightarrow\quad c_{\alpha} = \xi_{\alpha} \sqrt{\theta} + u_{\alpha},
\end{equation}
which is the continuous counterpart of Eq.~(\ref{eq:discrete_norm_peculiar}).
Both formulas correspond to a change of reference frame that leads to a Galilean and temperature invariant description of velocity distribution functions (resp. populations). In other words, by working with populations that propagate at a speed $c_{\alpha}$ (resp. $c_{i\alpha}$), macroscopic errors emerging from the approximation of the Maxwellian (e.g., Hermite polynomial expansion, numerical equilibrium, etc) boil down to zero, even for strong deviations from the reference state $(\bm{u},\theta)=(0,1)$.

In the discrete case, one can prove it by comparing the values of moments computed via discrete EDFs with those of the Maxwellian, for different values of the macroscopic velocity $u_{\alpha}$ and $\theta$ (see Section~\ref{subsubsec:numerical_EDF_operability_range}, Appendix~\ref{app:errorMacro} as well as Refs.~\cite{KORNREICH_PD_69_1993,FRAPOLLI_PhD_2017,HOSSEINI_PRE_100_2019,LATT_RSTA_378_2020}).
Hereafter, we will focus on an elegant alternative way to achieve the same goal in the continuous case by simply comparing the shape of the Maxwellian with its Hermite polynomial expansion. 
The latter relies on the fact that moments are quantitative measures related to the shape of a probability distribution function. Hence, distribution functions with a similar shape will have similar moments and vice versa.

With this aim in mind, let us first recall some basics about the polynomial expansion of the Maxwellian for a given reference frame. By definition, the Maxwell-Boltzmann EDF reads as
\begin{equation}
    f^{\mathrm{MB}}(\bm{x},\bm{\xi},t)=\dfrac{\rho}{(2\pi \theta)^{d/2}}\exp\bigg[-\dfrac{(\xi-u)^2}{2\theta}\bigg],
\end{equation}
and its Hermite polynomial expansion (truncated at order $N$) is expressed as~\cite{GRADb_CPAM_2_1949,SHAN_PRL_80_1998,SHAN_JFM_550_2006}
\begin{equation}\label{eq:hermiteExpansion}
    f^{\mathrm{MB},N}(\bm{x},\bm{\xi},t)= \dfrac{\rho}{\theta^{d/2}} w(\bm{\eta}) \sum_{n=0}^N \dfrac{1}{n!} a_{\mathrm{MB}}^{(n)}(\bm{x},t):\mathcal{H}^{(n)} (\bm{\eta}).
\end{equation}
$\mathcal{H}^{(n)}$ are Hermite polynomials at order $n$ computed via $\eta_{\alpha}$ ($\alpha=x$ or $y$ in 2D) which is a quantity yet to be defined.
$a_{\mathrm{MB}}^{(n)}$ are Hermite coefficients normalized by the density $\rho$, i.e.,
\begin{equation}
    a_{\mathrm{MB}}^{(n)}(\bm{x},t)=  \dfrac{1}{\rho} \int f^{\mathrm{MB}} (\bm{x},\bm{\xi},t) \mathcal{H}^{(n)}(\bm{\eta})\, \mathrm{d}\bm{\eta}.
\end{equation}
In Eq.~(\ref{eq:hermiteExpansion}), ``$:$'' further stands for Frobenius inner product also known as full tensor contraction, and $w$ is the weight function
\begin{equation}
    w(\bm{\eta}) = \dfrac{1}{(2\pi)^{d/2}}\exp\big(-\eta^2/2\big)
\end{equation}
used for the Gauss-Hermite quadrature. Several definitions of $\eta_{\alpha}$ exist depending on the reference frame considered~\cite{GUO_Book_2013}. 
For a reference frame at rest, $\eta_{\alpha}=\xi_{\alpha}$ and $a_{\mathrm{MB}}^{(n)}$  is a Hermite moment of the Maxwellian, whereas in the co-moving reference frame, $\eta_{\alpha}=\xi_{\alpha} - u_{\alpha}$ and $a_{\mathrm{MB}}^{(n)}$ is now a central Hermite moment~\cite{MATTILA_PF_29_2017,COREIXAS_PRE_100_2019,HOSSEINI_RSTA_378_2020}. 
In the most general case, the weight function further accounts for temperature variations through $\eta_{\alpha}=(\xi_{\alpha} - u_{\alpha})/\sqrt{\theta}$, hence leading to
\begin{equation}
    w = \dfrac{\theta^{d/2}}{\rho} f^{\mathrm{MB}}.
\end{equation}
%where the prefactor $1/\theta^{d/2}$ comes from the constraint $\int w = 1$.
Here, polynomial coefficients are central Hermite moments normalized by the temperature, and they satisfy the elegant mathematical property of all being null but the zeroth-order term~\cite{GRADb_CPAM_2_1949,SHAN_PRE_100_2019,LI_PRE_100_2019} -- central Hermite moments only share the same property in the isothermal case. 
Consequently, $a_{\mathrm{MB}}^{(n)}$ are velocity- and temperature-independent, whereas for $\eta_{\alpha}=\xi_{\alpha}-u_{\alpha}$, they will depend on temperature variations. 
For a reference frame at rest, these moments will finally depend on both velocity and temperature variations.

This behavior is illustrated in Fig.~\ref{fig:maxwell_vs_hermite}, where the Maxwellian and its three Hermite polynomial expansions are plotted in the 1D case. 
These polynomial expansions are done based on Hermite (static), central Hermite (adapt), and $\theta$-normalized central Hermite moments (adapt norm). 
All of them include terms up to $N=4$, which is the minimal configuration to recover the macroscopic behavior of the compressible Navier-Stokes-Fourier equations~\cite{GRADb_CPAM_2_1949,SHAN_PRL_80_1998,SHAN_JFM_550_2006}. 
As expected, the static configuration shows errors when the operating point is far from the reference state $(u,\theta)=(0,1)$. 
In addition, describing the Maxwellian in terms of central Hermite moments prevents any velocity-dependent mismatch, but it cannot correct errors induced by temperature variation. 
In the end, accounting for velocity and temperature variations in the propagation speed is the only way to obtain an error-free approximation of the Maxwellian, and consequently, the correct macroscopic behavior.

%Taking one step back and looking at already existing and present adaptive LBMs, it then seems clear that the cornerstone of these approaches is the description of populations in terms of velocity- and temperature-dependent lattices. The present interpolation-free formulation is then only proposed to derive 

%The reconstruction of missing information through various techniques --even though being of paramount importance to design efficient alternatives to Naviers-Stokes-Fourier solvers-- is only of practical interest and does not modify the macroscopic behavior of each of these adaptive LBMs.   
}}

\section{Linear stability analysis in a nutshell \label{app:VN}}
For non-linear systems of equations, this approach consists of introducing a perturbation into the linearized version of the considered system of equations in a fully periodic domain, then following its time evolution \cite{VONNEUMANN_Book_1966,HIRSCH_Book_2007}. 
In the context of LB solvers, under the assumption of a linear regime, one can approximate the field though a first-order Taylor-McLaurin expansion:
\begin{equation}
  f_i \approx \overline{f}_i + f'_i.
\end{equation}
Introducing this expansion into the target discrete system of equations one recovers the discrete linearized equations for the perturbation.

Given that the topic has been thoroughly treated in the literature~\cite{STERLING_JCP_123_1996,WORTHING_PRE_56_1997,LALLEMAND_PRE_61_2000,DELLAR_PRE_65_2002,ADHIKARI_PRE_78_2008,MARIE_JCP_228_2009,RICOT_JCP_228_2009,XU_JCP_230_2011,XU_JCP_231_2012,DUBOIS_CRM_343_2015,HOSSEINI_IJMPC_28_2017,CHAVEZMODENA_CF_172_2018,COREIXAS_PhD_2018,HOSSEINI_PRE_99_2019b,HOSSEINI_PRE_100_2019,WISSOCQ_JCP_380_2019,HOSSEINI_RSTA_378_2020,MASSET_JFM_897_2020,WISSOCQ_ARXIV_2020_07353,RENARD_ARXIV_2020_08477}, details and derivation of the final equations will be omitted. As an example, the linearization of the BGK-LBE leads to
\begin{equation}
  f'_i(\bm{x}+\bm{\xi}_i, t+1) = \bigg[\delta_{ij}\bigg(1-\frac{1}{\tau}\bigg) + \frac{1}{\tau}J_{ij}^{eq}\bigg] f'_j(\bm{x}, t).
\end{equation}
where $J_{ij}^{eq}=\partial f^{eq}_i/ \partial f_j$ is the Jacobian matrix of the EDF that is evaluated at $f_j=\overline{f}_j$.
For a detailed derivation of these Jacobians, interested readers are referred to \cite{CHAVEZMODENA_CF_172_2018,HOSSEINI_IJMPC_28_2017,HOSSEINI_PRE_99_2019b,HOSSEINI_PRE_100_2019,WISSOCQ_JCP_380_2019,HOSSEINI_RSTA_378_2020,WISSOCQ_ARXIV_2020_07353,RENARD_ARXIV_2020_08477}.\\
Introducing the standing waves (wave number $\bm{k}\in\mathbb{R}$ and time frequency $\omega \in\mathbb{C}$) into the linearized discrete time evolution equation
\begin{equation}
f'_j = F'_j \exp{\left(\mathrm{i} \omega t - \bm{k}\cdot\bm{x}\right)},
\end{equation}
one obtains a system of equations, through which the time-amplification factor of the perturbation [$\exp[\mathrm{i}\omega]$) can be obtained for different wave numbers $\bm{k}$. In order for the system to be linearly stable for the chosen set of parameters, {i.e.} velocity, temperature and non-dimensional viscosity, the real component of $\exp[\mathrm{i}\omega]$ must remain negative for all possible values of $\bm{k}$.

For all stability domains presented in Section~\ref{subsec:polynomial_EDF}, the 1D wave number space $k\in [0,\pi]$ is discretized using a resolution of 100 points , {i.e.} $\Delta k_x=\pi/100$. The 1D assumption is justified by the fact that in either 2D or 3D, more waves are evolving in the Fourier space, hence, leading to more (unstable) couplings between modes. Consequently, the stability domain of a 1D model can usually be considered as a good approximation for the upper limit of stability domains in either 2D or 3D~\cite{RENARD_ARXIV_2020_08477}. 
%end: To be modified

\section{Detailed description of numerical equilibria\label{app:numericalEq}}

\subsection{Exact formulation for polyatomic gases}

For LBMs to fully recover the macroscopic behavior of NSF equations, the discrete EDF must mimick up to the (trace of the) fourth-order moment of the Maxwellian. In the bidimensional case, it should then satisfy the following 13 constraints:
\begin{align}
    G_0 &= \sum_i f_i^{eq} - \rho, \label{eq:G0}\\
    G_{1,x} &= \sum_i f_i^{eq} \xi_{ix}- \rho u_{x},  \label{eq:G1x}\\
    G_{1,y} &= \sum_i f_i^{eq} \xi_{iy}- \rho u_{y}, \label{eq:G1y}\\
    G_{2,xx} &= \sum_i f_i^{eq} \xi_{ix}^2 - \rho (u_x^2 + T),  \label{eq:G2xx}\\
    G_{2,xy} &= \sum_i f_i^{eq} \xi_{ix}\xi_{iy} - \rho u_x u_y,  \label{eq:G2xy}\\
    G_{2,yy} &= \sum_i f_i^{eq} \xi_{iy}^2 - \rho (u_y^2 + T),  \label{eq:G2yy}\\
    G_{3,xxx} &= \sum_i f_i^{eq} \xi_{ix}^3 -  \rho u_x(u_x^2 + 3 T),  \label{eq:G3xxx}\\
    G_{3,xxy} &= \sum_i f_i^{eq} \xi_{ix}^2\xi_{iy} -  \rho (u_x^2 + T) u_y, \label{eq:G3xxy}\\
    G_{3,xyy} &= \sum_i f_i^{eq} \xi_{ix}\xi_{iy}^2 -  \rho u_x(u_y^2 + T), \label{eq:G3xyy}\\
    G_{3,yyy} &= \sum_i f_i^{eq} \xi_{iy}^3 -  \rho u_y(u_y^2 + 3 T), \label{eq:G3yyy}\\
    G_{4,xx\alpha\alpha} &= \sum_i f_i^{eq} \xi_{ix}^2\xi_{i\alpha}^2 - 2\rho[(E + 2T)u_x^2 + (E + T)T], \label{eq:G4xxaa}\\
    G_{4,xy\alpha\alpha} &= \sum_i f_i^{eq} \xi_{ix}\xi_{iy}\xi_{i\alpha}^2 - 2\rho[(E + 2T)u_x u_y], \label{eq:G4xyaa}\\
    G_{4,yy\alpha\alpha} &= \sum_i f_i^{eq} \xi_{iy}^2\xi_{i\alpha}^2 - 2\rho[(E + 2T)u_y^2 + (E + T)T], \label{eq:G4yyaa}
\end{align}
where the repetition of (Greek) indices stands for Einstein's summation rule on geometrical coordinates, e.g., $\xi_{i\alpha}^2=\xi_{ix}^2+\xi_{iy}^2$.

The above constraints~(\ref{eq:G0})-(\ref{eq:G4yyaa}) are then used for the computation of the numerical equilibrium
\begin{widetext}
\begin{align}
    f_i^{eq} &= \rho \exp\big[-\big(1 + \lambda_0 + \big\{\lambda_{1,x}\xi_{ix} 
    + \lambda_{1,y}\xi_{iy}\big\} 
    + \big\{\lambda_{2,xx}\xi_{ix}^2 
    + \lambda_{2,xy}\xi_{ix}\xi_{iy}
    + \lambda_{2,yy}\xi_{iy}^2\big\}\nonumber\\    
    &\quad\quad\quad\quad  + \big\{\lambda_{3,xxy}\xi_{ix}^2\xi_{iy}
    + \lambda_{3,xyy}\xi_{ix}\xi_{iy}^2
    + \lambda_{3,yyy}\xi_{iy}^3\big\}
    + \big\{\lambda_{4,xx}\xi_{ix}^2\xi_{i\alpha}^2
    + \lambda_{4,xy}\xi_{ix}\xi_{iy}\xi_{i\alpha}^2
    + \lambda_{4,yy}\xi_{iy}^2\xi_{i\alpha}^2\big\}
    \big)\big],\label{eq:feqNum_13Mom}
\end{align}
\end{widetext}
where $\lambda_n$ are Lagrange multipliers that are obtained from a root-finding algorithm as zeros of the constraints $G_n$.

In addition, it is important to understand that populations $f_i$ do not account for extra internal degrees of freedom (rotational and vibrational). Hence, the total energy used in the above constraints is defined as $2E = u_x^2+u_y^2 + DT$, and $D=2$. To further model polyatomic gases, one can rely on the double distribution function framework for which a second set of populations $g_i$ is used to impose the correct specific heat ratio $\gamma_r$. While populations $f_i$ transfer the monatomic information to $g_i$ through $g_i^{eq}$~(\ref{eq:gEq}), the feedback from $g_i$ to $f_i$ is implicit (i.e., no forcing term is used in the collision of $f_i$), and it occurs in the computation of the temperature
\begin{equation}\label{eq:polyTemp}
    T = \dfrac{1}{2\rho C_v}\bigg[\sum_i (\xi_{i\alpha}^2 f_i + g_i) - \rho u_{\alpha}^2\bigg]
\end{equation}
with $C_v=1/(\gamma_r-1)$ being the polyatomic heat capacity at constant volume. This ``polyatomic'' temperature is then injected in the constraints (\ref{eq:G0})-(\ref{eq:G4yyaa}) that are used for the computation of $f_i^{eq}$~(\ref{eq:feqNum_13Mom}), hence, closing the loop.

The latter methodology allows the user to compute $f_i^{eq}$ in a \textit{numerical} manner, i.e., even if the system (\ref{eq:G0})-(\ref{eq:G4yyaa}) does not have an analytical solution. 
The resulting LBM is then freed from the very constraining quadrature rules that impose a minimal lattice size to recover a macroscopic behavior of interest. 
Nevertheless, the user must cautiously choose the convergence criterion of the root-finding algorithm. 
If not, conservation issues will appear because the constraints will not be correctly imposed. In the end, it is sufficient to impose a convergence criterion of $10^{-12}$, which leads to constraint errors that oscillates around $10^{-14}$, i.e., close to machine precision.

\subsection{Reduced models}

For both DVMs and LBMs, numerical equilibria were first based on constraints corresponding to the conservation of mass, momentum, and energy. In the bidimensional case, this corresponds to the following set of \textit{four} constraints
\begin{align}
    G_0 &= \sum_i f_i^{eq} - \rho,\\
    G_{1,x} &= \sum_i f_i^{eq} \xi_{ix}- \rho u_{x},\\
    G_{1,y} &= \sum_i f_i^{eq} \xi_{iy}- \rho u_{y},\\
    G_{2,\alpha\alpha} &= \sum_i f_i^{eq}\xi_{i\alpha}^2 - 2\rho E.
\end{align}
The latter are then used to compute the numerical equilibrium 
\begin{equation}
    f_i^{eq} = \rho \exp\big[-\big(1 + \lambda_0 + \big\{\lambda_{1,x}\xi_{ix} 
    + \lambda_{1,y}\xi_{iy}\big\} 
    + \lambda_{2,\alpha\alpha}\xi_{i\alpha}^2
    \big)\big].\label{eq:feqNum_4Mom}
\end{equation}
Even if this methodology reduces the CPU cost --induced by the iterative computation of the equilibrium-- as compared to the more demanding 13-moment approach~(\ref{eq:13moments}), it also comes at the expense of part of physics since both convective and diffusive terms are not properly recovered (See Eq.~\ref{eq:NSF_13moments}). Usually, one can only recover the correct physics by increasing the size of the lattice, as pointed out in the parametric study conducted in Section 3(c) of our previous work~\cite{LATT_RSTA_378_2020}.

In the context of high-speed and high-Reynolds number flows, (numerical) errors related to diffusive phenomena have a lower impact on the accuracy of the solver than those related to convective phenomena. This is the reason why, NS solvers are usually based on second-order (centered) numerical schemes for diffusive fluxes whereas convective terms are discretized using higher-order schemes~\cite{HIRSCH_Book_2007,LEBRAS_AIAAJ_55_2017}. Applying the latter reasoning to numerical EDFs, one can improve the 4-moment methodology by accounting for all constraints related to convective fluxes --so that $\Delta_2=\Delta_3^{tr}=0$ in Eq.~(\ref{eq:NSF_13moments})-- which leads to correct results for compressible flow simulations in the low-viscosity regime~\cite{LATT_RSTA_378_2020}. The corresponding set of \textit{eight} constraints reads as
\begin{align}
    G_0 &= \sum_i f_i^{eq} - \rho,\label{eq:G0_8mom}\\
    G_{1,x} &= \sum_i f_i^{eq} \xi_{ix}- \rho u_{x},\label{eq:G1x_8mom}\\
    G_{1,y} &= \sum_i f_i^{eq} \xi_{iy}- \rho u_{y},\label{eq:G1y_8mom}\\
    G_{2,xx} &= \sum_i f_i^{eq} \xi_{ix}^2 - \rho (u_x^2 + T)\label{eq:G2xx_8mom},\\
    G_{2,xy} &= \sum_i f_i^{eq} \xi_{ix}\xi_{iy} - \rho u_x u_y\label{eq:G2xy_8mom},\\
    G_{2,yy} &= \sum_i f_i^{eq} \xi_{iy}^2 - \rho (u_y^2 + T),\label{eq:G2yy_8mom}\\
    G_{3,x\alpha\alpha} &= \sum_i f_i^{eq} \xi_{ix}\xi_{i\alpha}^2 -  2 \rho u_x(E + T),\label{eq:G3xaa_8mom}\\
    G_{3,y\alpha\alpha} &= \sum_i f_i^{eq} \xi_{iy}\xi_{i\alpha}^2 -  2 \rho u_y(E + T),\label{eq:G3yaa_8mom}
\end{align}
and it is used to compute the equilibrium
\begin{widetext}
\begin{equation}
    f_i^{eq} = \rho \exp\big[-\big(1 + \lambda_0 + \big\{\lambda_{1,x}\xi_{ix} 
    + \lambda_{1,y}\xi_{iy}\big\} 
    + \big\{\lambda_{2,xx}\xi_{ix}^2 
    + \lambda_{2,xy}\xi_{ix}\xi_{iy}
    + \lambda_{2,yy}\xi_{iy}^2\big\}
    + \big\{\lambda_{3,x\alpha\alpha}\xi_{ix}\xi_{i\alpha}^2
    + \lambda_{3,y\alpha\alpha}\xi_{iy}\xi_{i\alpha}^2\big\}
    \big)\big].\label{eq:feqNum_8Mom}
\end{equation}
%\end{widetext}

Several alternative formulations can be derived depending on the targeted physics, i.e., which error terms $\Delta_n$ and/or $\Delta_n^{tr}$ should be cancelled in Eq.~(\ref{eq:NSF_13moments}). 
As an example, if diffusive fluxes are negligible in the energy equation but not in the momentum equation, one could rely on the set of \textit{ten} constraints (\ref{eq:G0})-(\ref{eq:G3yyy}) so that only $\Delta_4^{tr}$ would be non-zero. 
One could further want to better fit the definition of 
``admissible'' spaces which requires that the maximal order of constraints should always be even~\cite{LEVERMORE_JSP_83_1996}. This is done adding the constraint
\begin{align}
    G_{4,\alpha\alpha\beta\beta} &= G_{4,xx\beta\beta} + G_{4,yy\beta\beta} \nonumber\\
    &= \sum_i f_i^{eq} \xi_{i\alpha}^2\xi_{i\beta}^2 - 2\rho[(E + 2T)(u_x^2+u_y^2) + 2(E + T)T], \label{eq:G4aabb}
\end{align}
to the set of either eight or ten constraints leading to 9- and 11-moment approaches [see Eqs.~(\ref{eq:9moments}) and~(\ref{eq:11moments}) respectively].

\section{Preliminary study on the choice of lattice and number of constraints\label{app:errorMacro}}

\textcolor{black}{To find a good trade off between accuracy, stability and efficiency, we followed the methodology introduced in our previous paper~\cite{LATT_RSTA_378_2020}. The latter consists in evaluating the stability domain and accuracy of numerical EDFs for several lattices. This is done by computing macroscopic deviations of the numerical EDF moments with respect to the Mach number, and at a given reference temperature.\\ 
This aspect was investigated in a preliminary study for lattices composed of 9, 13, 17, 21, 25 (zero-one-three or ``ZOT'' formulation) and 49 discrete velocities respectively (see Fig.~\ref{fig:lattices} and Tab.~\ref{tab:refTemp} for their reference temperature $T_0$). 
Deviations with respect to the Maxwellian moments are reported for a numerical EDF based on $M=4$ and $8$ constraints in Fig.~\ref{fig:errorMacro}. 
Results show that the D2Q49 lattice is by far the most accurate velocity discretization, but it is also the most computationally expensive. 
Macroscopic deviations grow quickly with the Mach number for 9, 17 and 25 discrete velocities, whatever the number of constraints chosen. Regarding the D2Q13 and D2Q21 lattices, most errors remain below or close to the $\varepsilon=10\%$ threshold, especially with the 8-moment approach. 
In summary, the D2Q21 lattice is less accurate than the D2Q49 lattice, but it is far more efficient both in terms of simulation time and memory storage. In addition, it shares similar accuracy properties with the D2Q13 lattice while being more stable (the root-finding solver converges up to $\mathrm{Ma}\approx 1.85$ for the D2Q21 lattice). For all these reasons, coupling the D2Q21 lattice with the 8-moment approach seems to be a good trade off in terms of accuracy, stability and efficiency.% But it is true that the D2Q13 lattice (based on the 8-moment approach) is also an interesting candidate for the simulation of compressible flows in the high-subsonic regime.}
}

\begin{figure*}[hp!]
\centering
\includegraphics[width=.25\textwidth]{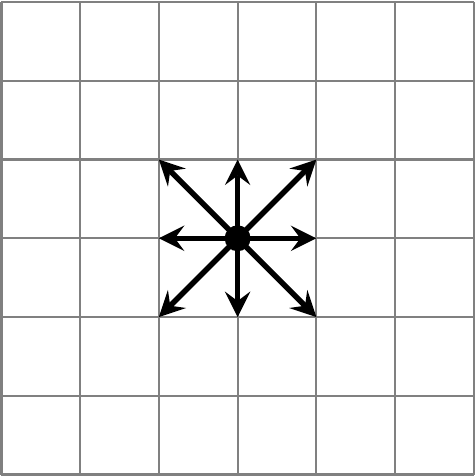}\quad
\includegraphics[width=.25\textwidth]{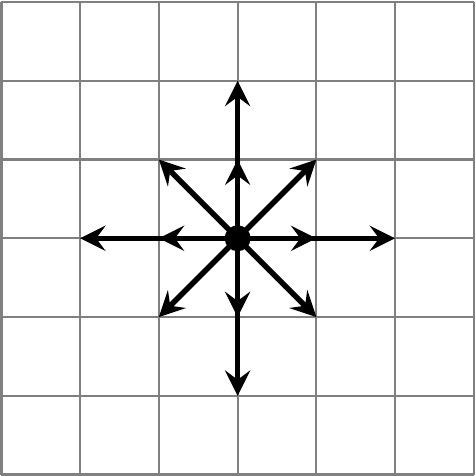}\quad
\includegraphics[width=.25\textwidth]{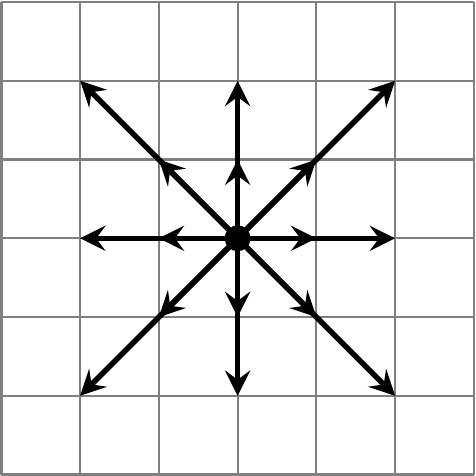}\\\vspace{0.25cm}
\includegraphics[width=.25\textwidth]{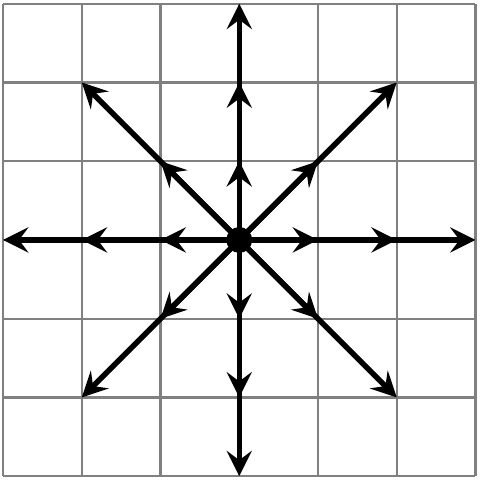}\quad
\includegraphics[width=.25\textwidth]{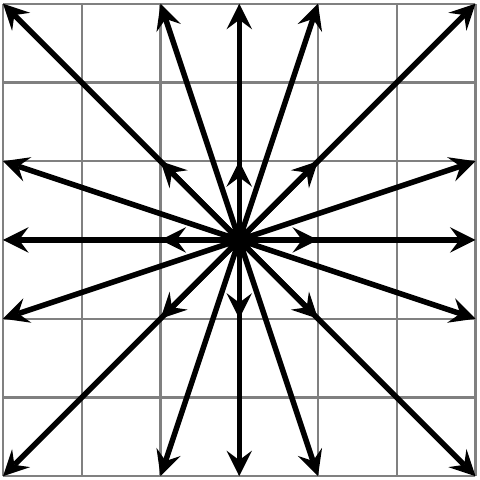}\quad
\includegraphics[width=.25\textwidth]{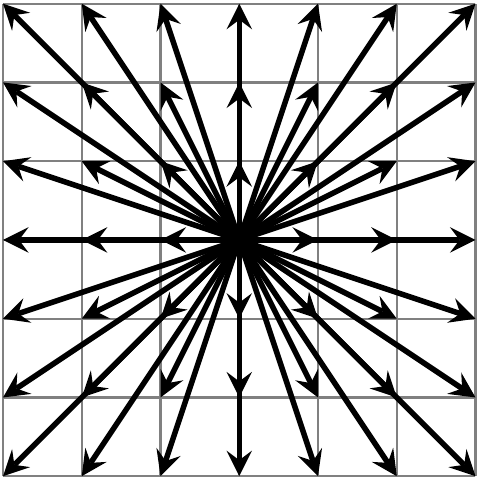}

\caption{\rcol{Velocity discretizations considered in the preliminary study. From top left to bottom right: D2Q9, D2Q13, D2Q17, D2Q21, D2Q25ZOT and D2Q49. Data are compiled from the works~\cite{PHILIPPI_PRE_73_2006,FRAPOLLI_PhD_2017} and therein references.}}
\label{fig:lattices}
\end{figure*}

\begin{table}[hb!]
\centering
\renewcommand{\arraystretch}{1.}
\renewcommand{\tabcolsep}{3mm}
\begin{tabular}{c c c c c c c}
\hline\hline
$\text{Lattice}$ & Q9 & Q13 & Q17 & Q21 & Q25ZOT & Q49 \\
$ T_0 $ & $1/3$ & $0.3744$ & $0.3764$ & $0.7$ & $0.5865$ & $0.6671$\\
\hline
\hline
\end{tabular}
\caption{\rcol{Reference temperatures $T_0$ obtained following instructions provided in our previous work~\cite{LATT_RSTA_378_2020}.}}
\label{tab:refTemp}
\end{table}

\begin{figure*}[btp]
\centering
\includegraphics[trim={0 0.25cm 0 0.25cm},clip,width=0.99\textwidth]{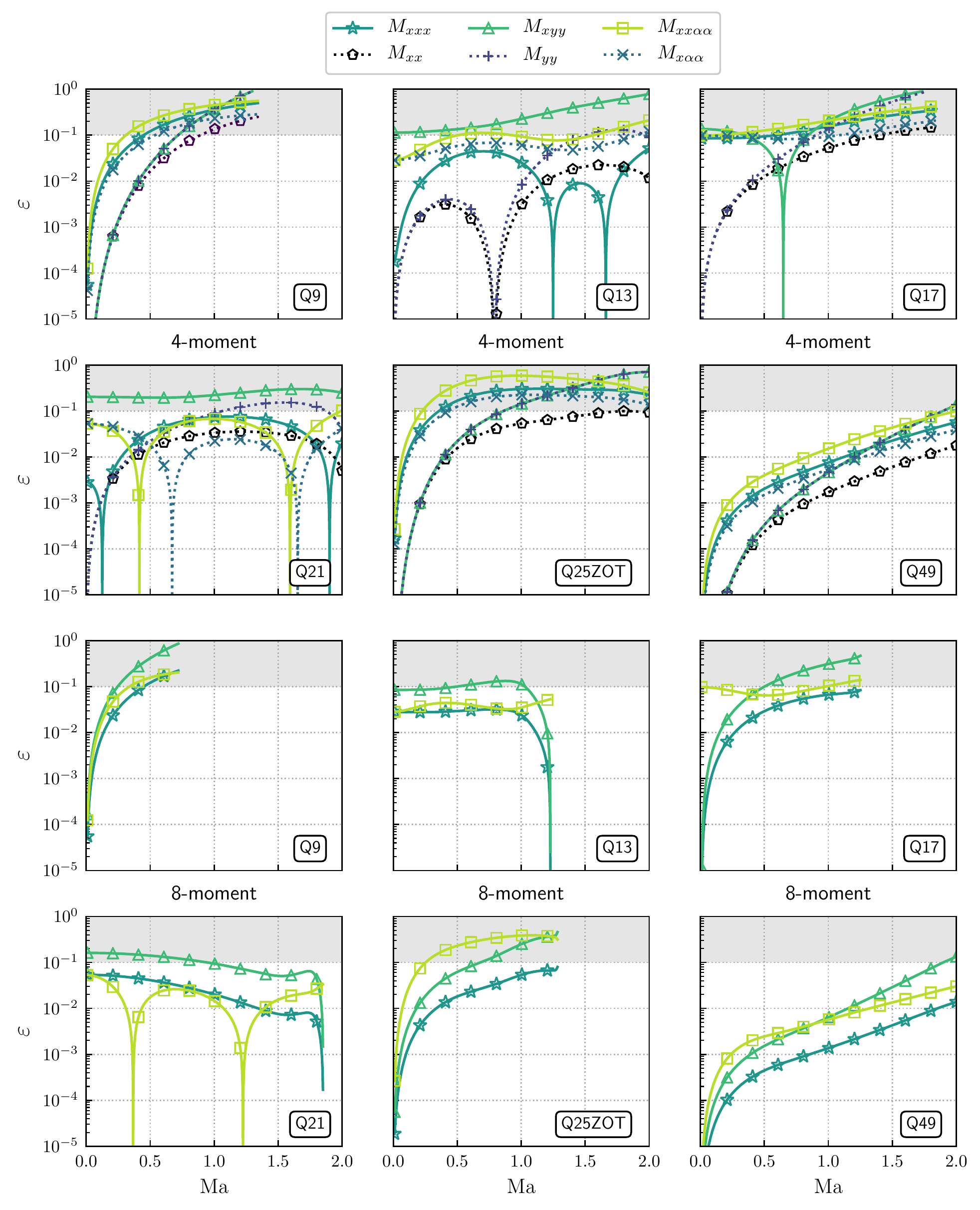}
\caption{\rcol{Mach number impact on macroscopic errors $\varepsilon$ for the 4-moment (top two rows) and 8-moment (bottom two rows) approaches. The flow propagates along the $x$-axis, and the grey zone starts at $\varepsilon = 10\%$. For the 8-moment approach, all second- and trace of third-order moments are enforced by the root-finding algorithm. Hence, corresponding errors are close to $10^{-15}$ (i.e., machine precision). For all configurations, $T=T_0$ with the reference temperature $T_0$ defined in Tab.~\ref{tab:refTemp} for each lattice.}}
\label{fig:errorMacro}    
\end{figure*}

\section{Flexible Prandtl number extension\label{app:PrandtlExtension}}
%\rcol{Say something about Grad reconstruction, Shakov's collision model and the fact that it is an extension of Shan's collision model.}

By relying on a single relaxation time approximation~(\ref{eq:coll}), the present DDF-LBM is restricted to $\mathrm{Pr}=1$. This can be corrected by adopting more sophisticated collisions models, e.g., the ES-BGK~\cite{HOLWAY_PoF_9_1966} or Shakov's~\cite{SHAKHOV_FD_3_1968} collision model. Both were introduced in the monatomic case, and extended to polyatomic gases by Andries et al.~\cite{ANDRIES_EJMBF_19_2000} and Rykov~\cite{RYKOV_FD_10_1975} respectively. In the following, we adopt Rykov's approach where we further suppose that collisions are elastic. This basically modifies the collision step~(\ref{eq:coll}) as follows:
\begin{alignat}{3}\label{eq:Shakhov}
    f_{i}^* &= f_i^{eq} &+ \bigg(1 - \dfrac{1}{\tau_f}\bigg) (f_{i}-f_{i}^{eq}) &+ \bigg(\dfrac{1}{\tau_f}-\dfrac{1}{\tau_{\mathrm{Pr}}}\bigg) \dfrac{q_{\alpha}\overline{\xi}_{i\alpha}}{\rho C_p T^2}\bigg(\dfrac{\overline{\xi}_i^2}{2T}-C_p\bigg)f_{i}^{eq},\\
    g_{i}^* &= g_i^{eq} &+ \bigg(1 - \dfrac{1}{\tau_g}\bigg) (g_{i}-g_{i}^{eq}) &+ \bigg(\dfrac{1}{\tau_g}-\dfrac{1}{\tau_{\mathrm{Pr}}}\bigg) \dfrac{2 q_{\alpha}\overline{\xi}_{i\alpha}}{\rho C_p}g_{i}^{eq},\label{eq:prandtlPoly}
\end{alignat}
with $\tau_f=\tau_g=0.5 + \nu/T$, and $\tau_{\mathrm{Pr}}=0.5 + (\nu/\mathrm{Pr})/T$. 

In order to extend the reconstruction of missing information at transition interfaces, one simply needs to replace the non-equilibrium contributions $h_i^{neq}=h_i-h_i^{eq}$ by $h_i^{(1),\mathrm{CE}}$ or
$h_i^{(1),\mathrm{Grad}}$ with $h_i=f_i$, $g_i$. Interestingly, when the reconstruction step is applied to all cells of the simulation domain (at each time step), one ends up with an extended regularized collision model, whose properties will be investigated in a future paper.
\end{widetext}

%\section{Decay of shear and thermal waves\label{app:decay}}

%\rcol{Hereafter, we further investigate the transport of thermal information in a \textit{viscous} medium. This is done simulating the dissipation of a 1D thermal wave, which basically corresponds to a (sinusoidal) thermal disturbance evolving in an uniform mean flow at constant pressure:
%\begin{equation}
%    T=T_0[1 + A \sin(2\pi x / L_x)], \: \rho = \rho_0[1 - A \sin(2\pi x / L_x)], 
%\end{equation}
%where all parameters are the same as for the entropy spot. In the linear regime, this wave amplitude decays following an exponential law: $\exp[-\nu_{\mathrm{th}}t]$, where $\nu_{\mathrm{th}}=\nu/\mathrm{Pr}$ is the thermal diffusivity coefficient. 
%}

%\nocite{*}
%\bibliography{biblio}% Produces the bibliography via BibTeX.

\providecommand{\noopsort}[1]{}\providecommand{\singleletter}[1]{#1}%

\end{document}